\documentclass[12pt,preprint]{aastex}

\usepackage{subfigmat}
\usepackage{lscape}

\newcommand{\figname}{Figure~}
\newcommand{\tabname}{Table~}
\newcommand{\eqnname}{Eqn.~}
\newcommand{\sdssg}{$g$}
\newcommand{\sdssr}{$r$}
\newcommand{\sdssi}{$i$}
\newcommand{\ursdss}{$u$ - $r$}
\newcommand{\chisq}{$\chi^{2}$}
\newcommand{\gsim}{\stackrel{>}{{}_{\sim}}}

\newcommand{\reff}{$r_{\rm eff}$}
\newcommand{\reffm}{r_{\rm eff}}
\newcommand{\rfiber}{$r_{\rm fiber}$}
\newcommand{\Halpha}{H$\alpha$}
\newcommand{\Hbeta}{H$\beta$}
\newcommand{\Hgamma}{H$\gamma$}
\newcommand{\Hdelta}{H$\delta$}
\newcommand{\Heta}{H$\eta$}
\newcommand{\Htheta}{H$\theta$}
\newcommand{\Halpham}{\textrm{H}\alpha}
\newcommand{\Hbetam}{\textrm{H}\beta}
\newcommand{\balmerdec}{H$\alpha$/H$\beta$}
\newcommand{\ba}{b/a}
\newcommand{\tauxfaceon}{\tau_{x}(1)}
\newcommand{\tauxinclin}{\tau_{x}(\ba)}

\newcommand{\Mrpetro}{$M_r$}
\newcommand{\colorexcessbalmer}{{E(B-V)}_{\rm Balmer}}
\newcommand{\colorexcessstar}{{E(B-V)}_{\rm stars}}
\newcommand{\BVstar}{{(B-V)}_{\rm stars}}
\newcommand{\balmertau}{\tau^{l}_{\rm B}}
\newcommand{\kms}{km~s$^{-1}$}
\newcommand{\wavenumber}{\tilde{\nu}}
\newcommand{\balmerdecHIItheoretical}{2.86}
\newcommand{\redshift}{$z$}
\newcommand{\fracdevr}{fracDeV$_r$}
\newcommand{\axisratio}{$b/a$}
\newcommand{\sdssmrlimit}{$17.77$~mag}
\newcommand{\psamplez}{$0 - 0.2$}
\newcommand{\psamplefracdevr}{$0 - 1$}
\newcommand{\samplefracdevr}{$0 - 0.1$}
\newcommand{\samplemagnitude}{-19.5 to -22~mag}
\newcommand{\sampleredshift}{$0.065 - 0.075$}
\newcommand{\sampleoneredshift}{$0.03 - 0.04$}
\newcommand{\sampletworedshift}{$0.065 - 0.075$}
\newcommand{\samplethreeredshift}{$0.1 - 0.11$}
\newcommand{\redsequence}{\ursdss \ $> 2.4$}
\newcommand{\redsequencecontamination}{$\approx~5$\%}

\newcommand{\baedgeonrange}{0.1 - 0.2}
\newcommand{\baedgeonextra}{0.1}

\newcommand{\bafaceonrange}{0.9 - 1.0}
\newcommand{\badrop}{$0.0 - 1.0$}
\newcommand{\betar}{0.26} 
\newcommand{\betarsd}{0.01} 
\newcommand{\rfibervsreffintrinsic}{0.45} 
\newcommand{\rextg}{1.2~mag} 
\newcommand{\rextginclin}{$\approx{0.4}$~mag} 
\newcommand{\rextrinclin}{$\approx{0.2}$~mag} 
\newcommand{\rextiinclin}{$\approx{0.1}$~mag} 
\newcommand{\ebvstartobalmer}{0.23 \pm 0.20}
\newcommand{\extgfaceon}{0.2~mag}
\newcommand{\ebvmodeldependency}{0.1~unit}
\newcommand{\modelofchoice}{slab model}
\newcommand{\balmerdecPercentileVSBayesian}{1.5}
\newcommand{\waveS}{3700}
\newcommand{\waveE}{8000}
\newcommand{\waveMid}{5000}
\newcommand{\onesigma}{1$\sigma$}
\newcommand{\nsigmaline}{2$\sigma$}
\newcommand{\restEWMin}{0.7~\AA}
\newcommand{\stellarabs}{1.3~\AA}

\begin{document}

\title{Extinction in Star-Forming Disk Galaxies from Inclination-Dependent Composite Spectra}

\author{Ching-Wa Yip\altaffilmark{1},  Alex S. Szalay\altaffilmark{1},
  Rosemary            F.~G.~Wyse\altaffilmark{1},           L\'aszl\'o
  Dobos\altaffilmark{2},  Tam\'as  Budav\'ari\altaffilmark{1},  Istvan
  Csabai\altaffilmark{2}}

\altaffiltext{1}{Department  of  Physics   and  Astronomy,  The  Johns
  Hopkins      University,     Baltimore,     MD      21218,     USA.}
  \altaffiltext{2}{Department   of   Physics   of   Complex   Systems,
  E\"{o}tv\"{o}s Lor\'and University, H-1117 Budapest, Hungary.}

					     \email{cwyip@pha.jhu.edu;
					     szalay@pha.jhu.edu}

\begin{abstract}
Extinction  in   galaxies  affects  their   observed  properties.   In
scenarios describing the distribution  of dust and stars in individual
disk galaxies, the amplitude of the extinction can be modulated by the
inclination  of  the  galaxies.   In  this  work  we  investigate  the
inclination  dependency  in  composite  spectra of  star-forming  disk
galaxies  from the  Sloan  Digital Sky  Survey  Data~Release~5.  In  a
volume-limited sample within a  redshift range \sampleredshift \ and a
\sdssr-band Petrosian absolute magnitude range \samplemagnitude \ which
exhibits a flat distribution  of inclination, the inclined relative to
face-on extinction  in the stellar  continuum is found  empirically to
increase with inclination in the  \sdssg, \sdssr \ and \sdssi \ bands.
Within the central $0.5$  intrinsic half-light radius of the galaxies,
the \sdssg-band  relative extinction in the stellar  continuum for the
highly-inclined  objects  (axis  ratio  $\ba =  $  \baedgeonextra)  is
\rextg, agreeing  with previous studies.  The extinction  curve of the
disk  galaxies   is  given   in  the  restframe   wavelengths  $\waveS
-\waveE$~\AA,  identified with major  optical emission  and absorption
lines  in  diagnostics.   The  Balmer decrement,  \balmerdec,  remains
constant  with  inclination,  suggesting  a  different  kind  of  dust
configuration and/or reddening mechanism in the \ion{H}{2} region from
that in the stellar continuum.  One factor is shown to be the presence
of spatially non-uniform interstellar extinction, presumably caused by
clumped dust in the vicinity of the \ion{H}{2} region.
\end{abstract}


\section{Introduction} \label{section:intro}

Extinction in galaxies affects  their observed properties.  Ideally it
should be  taken into account when inferring  the intrinsic properties
of  the galaxies.  One  empirical approach  in studying  extinction in
disk galaxies  is to compare  the observed properties of  objects with
similar     intrinsic    properties    at     different    inclination
\citep{1958MeLu2.136....1H}.  Optically  thin galaxies, when inclined,
should show higher  surface-brightness than their face-on counterparts
because of  a larger column of  stars observed in  a smaller projected
area on the sky. This  approach has been adopted into various analyses
by  many authors  to study  extinction in  disk  galaxies \citep[e.g.,
][]{1989MNRAS.239..939D,1990Natur.346..153V,1992MNRAS.254..677H,1993MNRAS.260..491D,1994AJ....107.2036G,1994A&A...283...12B,1995osd..conf.....D}.

Recently,  there are  revived  interests on  the  topic of  extinction
because of the large sample  of galaxies made available by various sky
surveys     including     the     Sloan     Digital     Sky     Survey
\citep[SDSS;][]{2000AJ....120.1579Y}.   Progress   has  been  made  in
studying extinction  in galaxies empirically by  using techniques that
are  depend on  or  independent of  the inclination-related  approach.
\citet{2007ApJ...659.1159S} demonstrated the inclination dependency of
dust extinction in disk-dominated galaxies from the SDSS, by using the
$z$-band extinction\footnote{The $z$-band  extinction was estimated by
\citet{2003MNRAS.341...33K} who used the difference between the colors
from  the  stellar population  model  and  from  the measurement,  and
extrapolation to the $z$-band  assuming a standard extinction curve.},
and  an inclination  measure being  the apparent  major to  minor axes
ratio.  \citet{2007AJ....134.2385H} applied the technique of occulting
galaxy  pairs  \citep{1992Natur.359..129W}  to  83  SDSS  spiral  disk
galaxies in  in probing  the opacity of  the foreground  galaxies, and
found that the optical depths ($\tau$) in the \sdssg, \sdssr, \sdssi \
bands are  typically within the range  $0 - 1.5$ across  the disk from
approximately  ${0.5}$  to  ${4}$~half-light  radii (\reff,  in  their
\figname6).   \citet{2008ApJ...687..976U}   looked  at  disk-dominated
galaxies  from the  SDSS and  found that  the highly  inclined objects
($\ba   \sim{\baedgeonextra}$)    exhibit   $\sim{1}$~mag   extinction
correction      relative      to      the      face-on      magnitude.
\citet{2009ApJ...691..394M}    studied   the    observed   photometric
properties  of the  SDSS galaxies  as a  function of  inclination, and
concluded that the  inclination-dependent extinction correction in the
\sdssg \ band could reach $\sim{1}$~mag with a median value of 0.3~mag
for  a sample of  disk galaxies.   \citet{2008ApJ...681..225B} applied
the  inclination-corrected   magnitudes  to  the   context  of  galaxy
classification.  \citet{2007MNRAS.379.1022D} developed a sophisticated
approach to derive iteratively the extinction-inclination relation for
both the  bulge and disk components  of disk galaxies,  and found that
the edge-on  to face-on  relative extinction is  $\approx~{2}$~mag and
$\approx~{1}$~mag, respectively  for the bulge and the  disk.  This is
an unique approach to date to simultaneously perform the bulge-to-disk
decomposition and derive  the extinction-inclination relation for both
components.

Inspired by these  studies, and that the SDSS provides  for us a large
sample  of galaxy  spectra  within a  fixed central  3\arcsec-diameter
fiber, and that the calibration  of the spectra is independent of that
in  the  SDSS photometric  data  used by  \citet{2007ApJ...659.1159S},
\citet{2007AJ....134.2385H},      \citet{2008ApJ...687..976U}      and
\citet{2009ApJ...691..394M},  we  now  investigate  if  there  is  any
inclination dependency in the observed galaxy spectra; and what is its
wavelength dependency, e.g., the effect in the gaseous component, such
as the \ion{H}{2}  region, vs. that in the  stellar continuum; and the
comparison of  extinctions among different elements.   Our approach is
to  explore  the  inclination   dependency  of  composite  spectra  of
star-forming  disk galaxies  in a  well defined  volume-limited sample
that  exhibits a  flat distribution  of the  galaxy  inclination.  The
volume-limited   nature  of  the   sample  eliminates   the  Malmquist
bias\footnote{In that, one preferentially draws galaxies of higher (or
lower) luminosity  at higher  (or lower) redshift,  for example,  in a
flux-limited survey like  the SDSS.}  in the analysis.   The sample is
also selected to exhibit a flat distribution of the inclination angle,
so that at a chosen redshift  ($z$) range, a galaxy of luminosity in a
defined range would be observed by the survey, {\it independent} of
the inclination  angle (or other properties)  of the galaxy.   It is a
very  important step  because one  can then  assume randomness  in the
intrinsic  properties of  the  sampled galaxies.   We further  confine
ourselves  to  a  sample of  a  narrow  redshift  range, so  that  any
difference detected in the observed properties between the two
composite spectra at different inclination  would not be mainly due to
the difference in  the redshifts of the galaxies.   On the other hand,
the composite spectra,  or the sample-average spectra, are  used in an
effort to  average the  variation in the  intrinsic properties  of the
galaxies, as  the actual optical depth of  individual galaxies depends
on  detailed properties  such as  the distribution  of dust,  gas, and
stars,  the structure of  the galaxies  such as  the presence  of arms
\citep[e.g.][]{1980ApJS...43...37E,1996ApJ...467..175B},   the  Hubble
type, etc..

We introduce  the galaxy  sample in \S\ref{section:data}.   We present
results  on the  inclination dependency  of composite  spectra  of the
galaxies  in  \S\ref{section:inclination}.   We derive  the  empirical
continuum extinction  in the  SDSS \sdssg, \sdssr,  \sdssi \  bands in
\S\ref{section:extinction},  the  model-based  face-on  extinction  in
\S\ref{section:theoreticalmodel},   and  the  extinction   curve  (the
extinction    as   a    function   of    restframe    wavelength)   in
\S\ref{section:extcurve}.   We study the  relation between  the Balmer
decrement  and  the   inclination-dependent  continuum  extinction  in
\S\ref{section:balmerdec}.     We    summarize    the    results    in
\S\ref{section:discussion}.

We use the term ``reddening'' to refer to the reduction in flux in the
shorter wavelengths  relative to longer wavelengths, or  a decrease in
the    steepness     of    the    spectral     slope.     We    follow
\citet{1995osd..conf.....D} and use  the terminology ``extinction'' to
refer  to  the  {\it  reduction}  of flux  due  to  absorption  and/or
scattering  by  dust.  We  use  ``attenuation''  interchangeably  with
extinction.  Following  the convention  in  the  SDSS  we express  the
spectra in vacuum wavelengths.

\section{The Sample} \label{section:data}

As part  of the  Sloan Digital Sky  Survey \citep{2000AJ....120.1579Y}
spectra are  taken with fibers of 3\arcsec  \, diameter (corresponding
to  0.18~mm at  the focal  plane for  the 2.5~m,  f/5  telescope). All
sources are  selected from  an initial imaging  survey using  the SDSS
camera  described  in   \citet{1998AJ....116.3040G}  with  the  filter
response curves as described in \citet{1996AJ....111.1748F}, and using
the imaging  processing pipeline of  \citet{2001ASPC..238..269L}.  The
astrometric  calibration is described  in \citet{2003AJ....125.1559P}.
The   photometric   system    and   calibration   are   described   in
\citep{2002AJ....123.2121S,2001AJ....122.2129H,2004AN....325..583I}.
The  selection criteria  of  the ``Main''  galaxies for  spectroscopic
follow-up observations are discussed by \citet{2002AJ....124.1810S}.

\begin{figure}
\epsscale{1.0}\plottwo{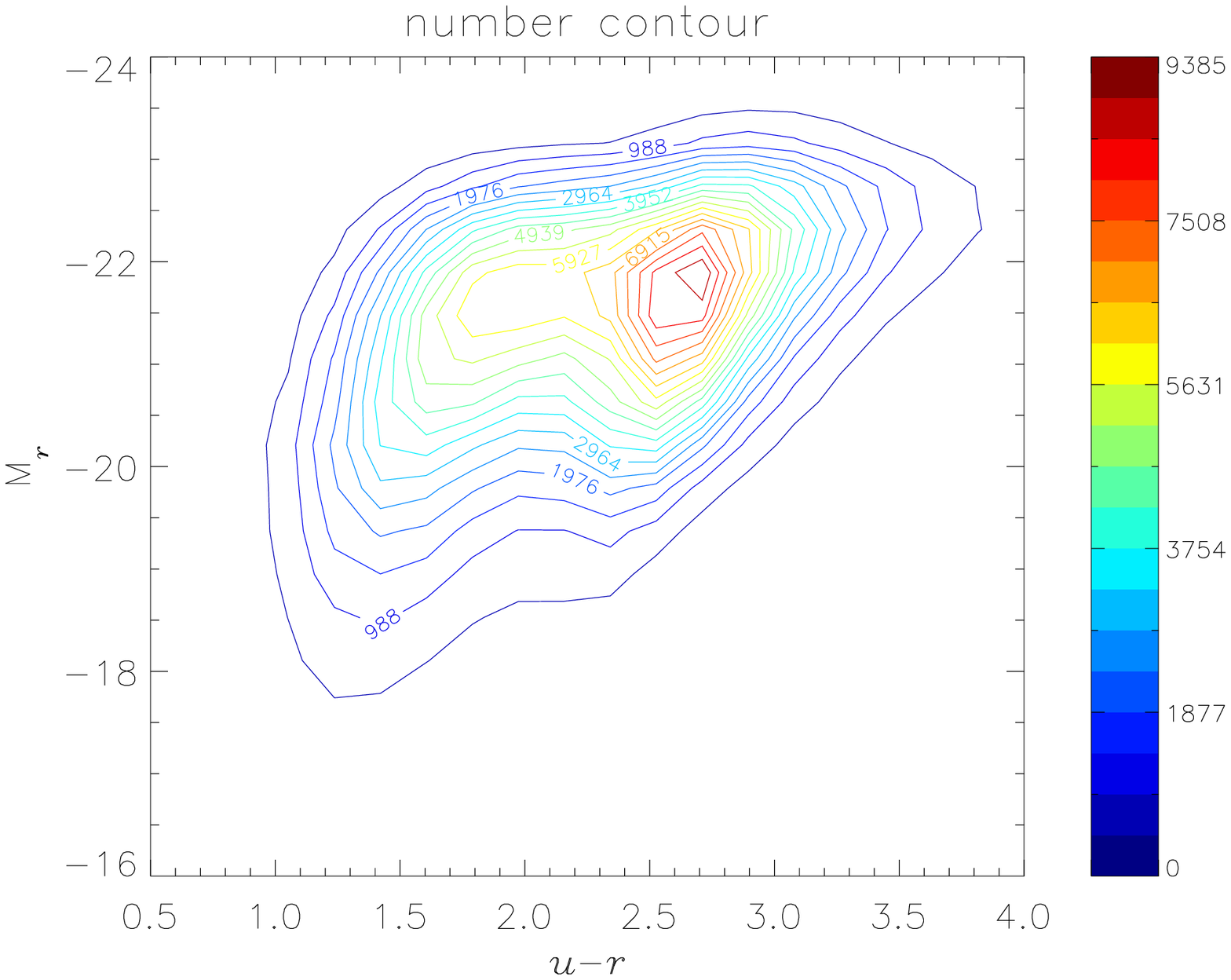}{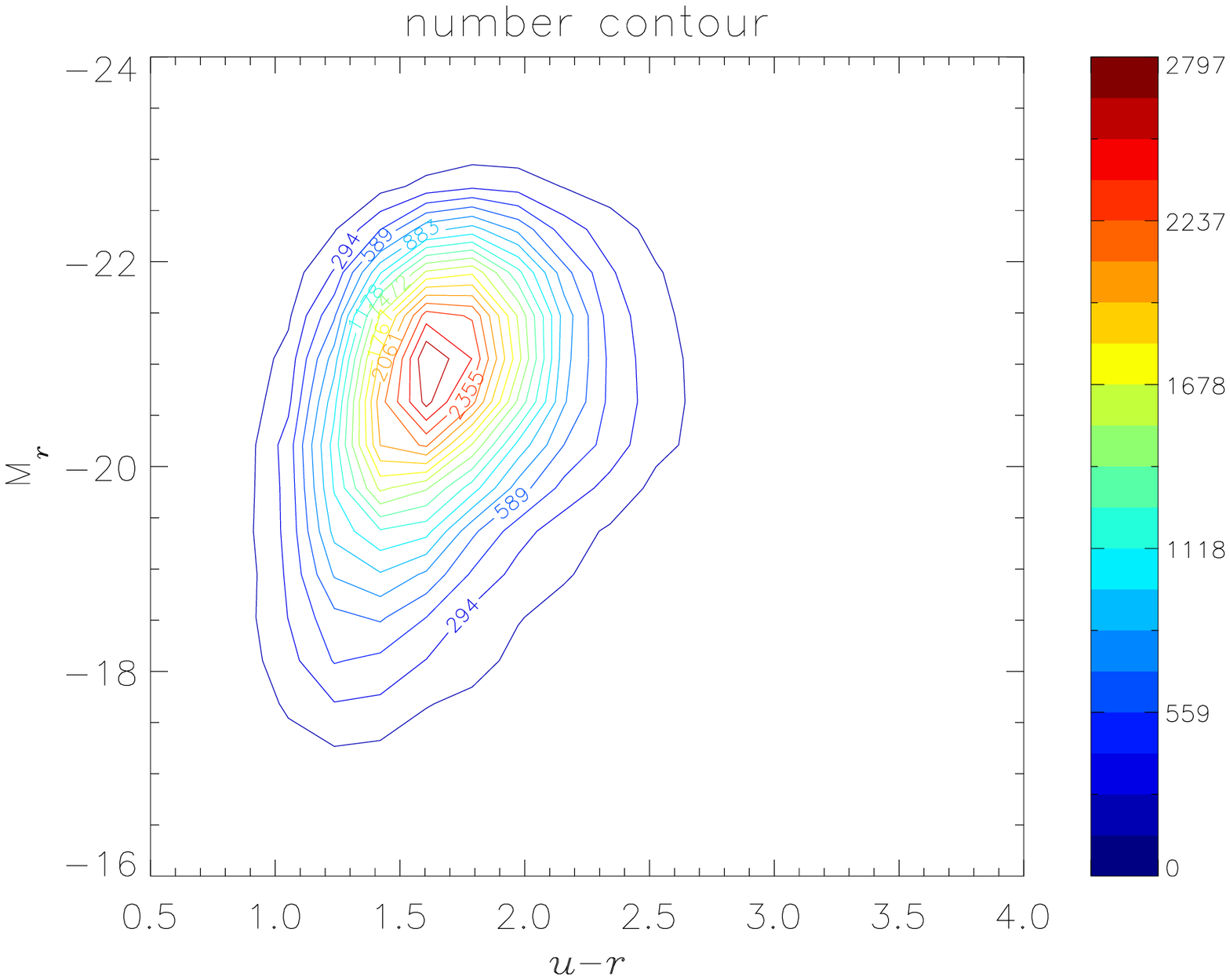}
\caption{Left: the restframe $r$-band Petrosian absolute magnitude vs.
the restframe \ursdss  \ color for the galaxies  in the parent sample.
The range of \fracdevr \  is from \psamplefracdevr. Both the magnitude
and the color  are K-corrected.  Right: same as  the left, but plotted
with  only  the   galaxies  that  lie  in  a   \fracdevr  \  range  of
\samplefracdevr.  Compared with the left figure, now the blue galaxies
are sampled predominantly, with a contamination by the red sequence of
\redsequencecontamination.}
\label{fig:psample1_Mr_u_r_sdss}
\end{figure}

Our sample is taken from  the spectra of star-forming disk galaxies in
the  SDSS  Data Release  5  (DR5).   We  determine the  criterion  for
selecting   disk-dominated  galaxies   with  the   approach   used  in
\citet{2008ApJ...687..976U},  who  characterized  that  galaxies  with
exponential radial  light profiles  fall on the  blue sequence  in the
color-magnitude diagram.  Based on  the full DR5 spectroscopic sample,
we define  a {\it parent  sample} of galaxies at  redshifts \psamplez,
which covers the median redshift  of the Main galaxy sample, about 0.1
\citep{2002AJ....124.1810S}.  The  galaxies in the  parent sample have
\fracdevr \ = \psamplefracdevr \ in the \sdssr \ band, where \fracdevr
\ equals  $0$ for  a pure exponential,  and equals  $1$ for a  pure de
Vaucouleurs  light profile as  provided by  the SDSS  imaging pipeline
\citep{2001ASPC..238..269L}.   The   color-magnitude  diagram  of  the
parent      sample      is     shown      in      the     left      of
\figname\ref{fig:psample1_Mr_u_r_sdss},  in the  \sdssr-band Petrosian
absolute  magnitude,  \Mrpetro,   vs.   \ursdss.   Both  the  absolute
magnitudes  and  the colors  are  K-corrected,  and corrected  against
foreground     dust     using    the     Milky     Way    dust     map
\citep{1998ApJ...500..525S}.   Two main distributions  are discernible
in the  color-magnitude space,  corresponding respectively to  the red
sequence (\redsequence)  and the blue  sequence.  Based on  the parent
sample, we determine the appropriate \fracdevr \ cut for selecting the
blue  galaxies  by repeatedly  narrowing  the  \fracdevr  \ range  and
inspecting  the  resultant  color-magnitude diagrams.   Galaxies  with
\fracdevr \ in the range of \samplefracdevr \ are fairly smooth in the
color-magnitude        distribution        (the        right        of
\figname\ref{fig:psample1_Mr_u_r_sdss}),  with a contamination  by the
red galaxies  of \redsequencecontamination \ (number ratio  of the red
sequence   to  the   total).   The   galaxies  with   \fracdevr   \  =
\samplefracdevr \ are used in the subsequent analysis.

\begin{figure}
\epsscale{0.8}\plotone{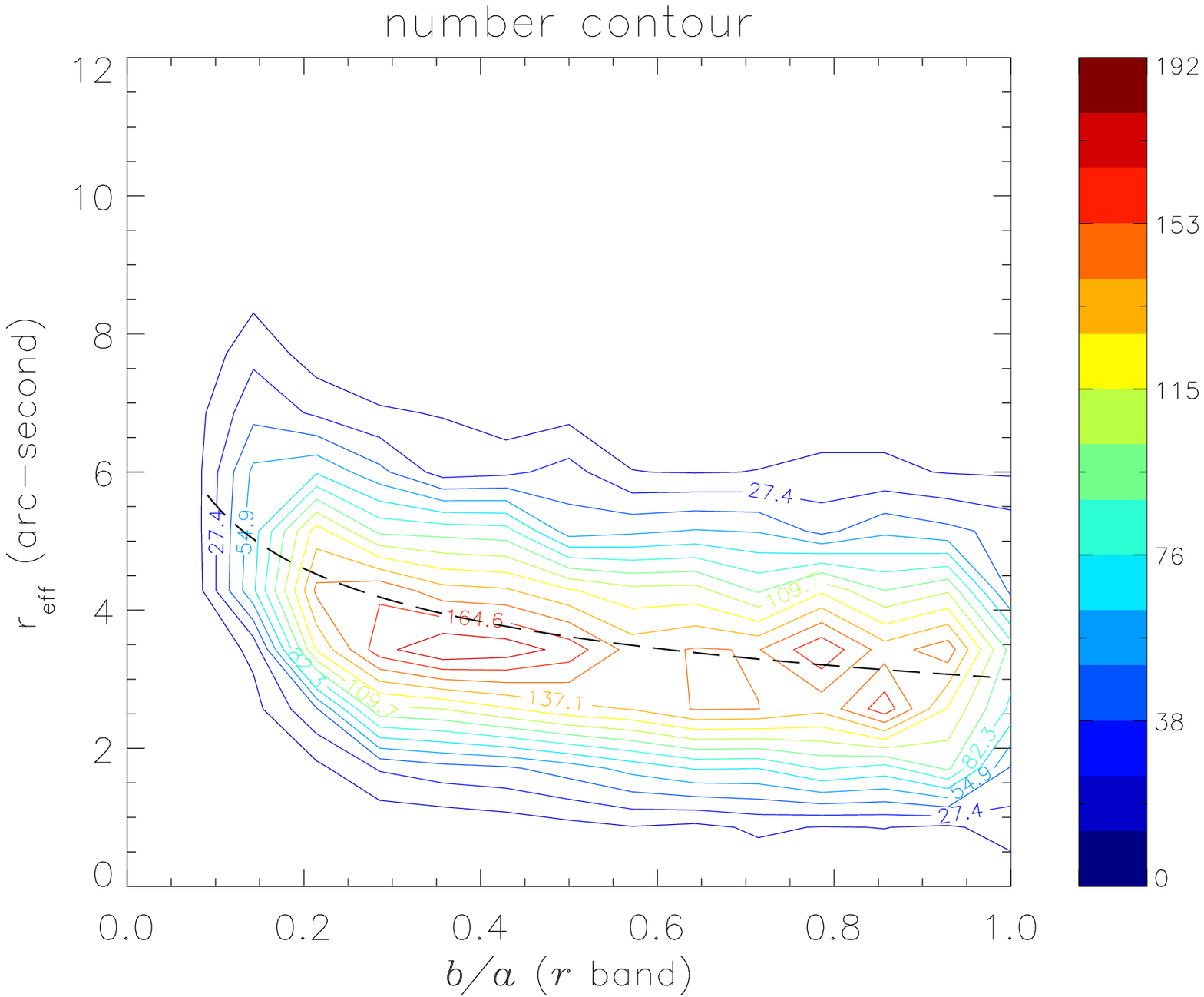}
\caption{The \sdssr-band apparent half-light radius, ${\reffm}^{\ba}$,
  as a  function of inclination of  the galaxies in  the parent sample
  that lie in a \fracdevr  \ range \ from \samplefracdevr.  The dashed
  line   is   the  best-fit   functional   form  ${\log}_{10}   \left(
  {\reffm}^{\ba} \right)  = {\log}_{10} \left(  {\reffm}^{1} \right) -
  \betar \, {\log}_{10} \left( \ba \right)$.}
\label{fig:psample1_expradr_vs_expabr_vlim_Mr_sdss_fracdevr_cut}
\end{figure}

  The inclination of a galaxy  is represented by the ratio between the
minor and major axes in the \sdssr \ band, expab$_r$ (or $\ba$ in this
paper for simplicity) provided by the SDSS through fitting each galaxy
image  with an exponential  light profile,  arbitrary axis  ration and
position angles.  Apparent axis ratio is shown to be a good measure of
the  average  intrinsic  inclination   of  a  sample  of  galaxies  by
\citet{2007ApJ...659.1159S},   who    used   the   approach    as   in
\citet{2004ApJ...601..214R}  to   simulate  the  average  relationship
between  the  intrinsic  and  apparent  inclinations in  a  sample  of
galaxies   of  Gaussian-distributed  intrinsic   thickness,  intrinsic
ellipticity that are observed at different viewing angles.\footnote{We
note that,  however, the galaxy samples they  considered are different
from   ours,  so   that  \citet{2007ApJ...659.1159S}   used  simulated
disk-only     galaxies,    and     \citet{2004ApJ...601..214R}    used
disk-dominated galaxies from the SDSS  at similar redshift range to us
but  larger  apparent  half-light  radius,  $\gsim{5}$\arcsec  \  (see
\figname\ref{fig:psample1_expradr_vs_expabr_vlim_Mr_sdss_fracdevr_cut}
for the size of our galaxies).}

\subsection{Volume-Limited Sample} \label{section:volumelimited}

Based  on  the  parent  sample   with  a  constrained  \fracdevr  \  =
\samplefracdevr,  several volume-limited  samples  are constructed  by
selecting galaxies in  narrow ranges of redshift.  From  these we seek
sample(s) that exhibit a  flat distribution of the galaxy inclination.
We emphasize  that this approach  is purely empirical but  is suitable
for  our  purpose,  as  such  a  volume-limited  sample  with  a  flat
distribution of galaxy inclination is  obtained. By the nature of this
approach, we neglect the step to {\it correct} for any possible biases
that are likely  to exist in the galaxy sample.   Rather, we resort to
the cancellation  of all of the  possible biases that gives  rise to a
flat distribution  of inclination.  In  the current context,  a biased
sample  is the  one which  preferentially includes  more  face-on than
edge-on galaxies,  or vice versa.   We summarize three major  kinds of
biases:

\begin{itemize} 
\item[1.] The flux-limited nature of the SDSS -- the flux limit in the
  SDSS spectra forbids the inclusion of inclined galaxies that are too
  dim to be observed at certain redshift.
\item[2.]  The use  of observed properties in sample  selection -- the
observed properties  are presumably  dust-attenuated.  The use  of the
inclination-induced reddened colors would cause a bias against edge-on
galaxies.  Besides,  the bulge-to-total  flux ratio (B/T)  of inclined
galaxies was found to decrease \citep{2004A&A...419..821T} because the
bulge is more  attenuated than the disk, implying  the use of observed
\fracdevr \ (correlated  with B/T) would cause a  bias towards edge-on
galaxies.
\item[3.]  The  bias that  is tough to  be controlled  without further
simulation  --  both the  intrinsic  thickness  of  galaxies, and  the
seeing, would make edge-on galaxies to appear rounder.
\end{itemize}

\noindent
The  sample   selection  for  studying   inclination-dependent  galaxy
properties is therefore a highly complex issue. This is exactly why we
seek an  empirical approach  as a pilot  study. Naturally  future work
should   be  focused  on   eliminating  these   biases  in   the  SDSS
spectroscopic  samples that  are  used for  inclination studies.   One
possible  approach is  to derive  all  of the  observed properties  of
interest that  are subjected to  both the inclination effects  and the
selection biases, through empirical relations in the literature and/or
simulation.   We  may then  work  backwards  to  derive the  intrinsic
properties of these galaxies.

\begin{figure}
\begin{subfigmatrix}{3}
     \centering \subfigure[\redshift \ =  \sampleoneredshift.]{
          \label{fig:z1}
          \includegraphics[width=.35\textwidth,angle=0]{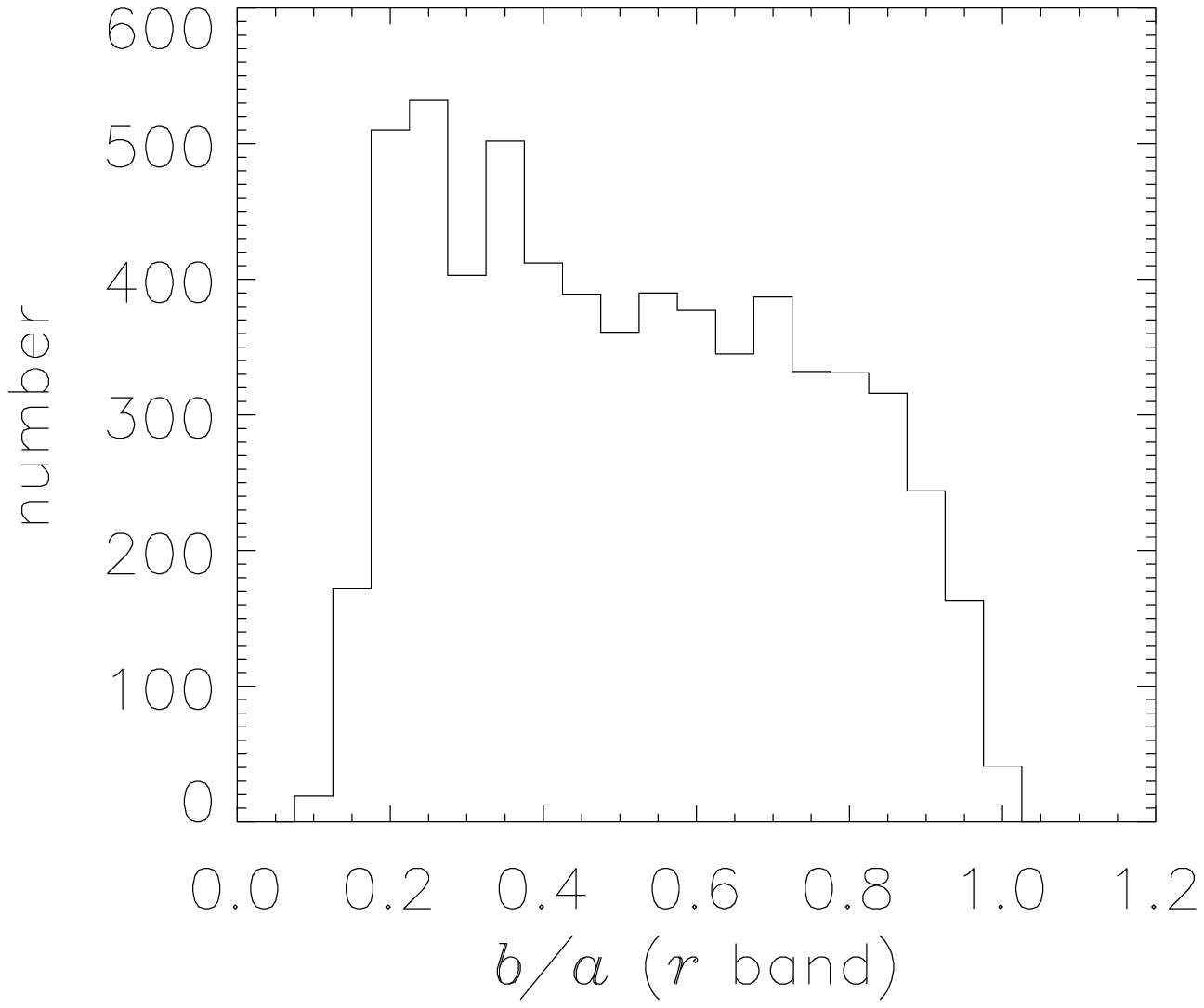}}
     \hspace{-1.5in}
     \subfigure[\redshift \ = \sampletworedshift.]{
          \label{fig:z2}
          \includegraphics[width=.35\textwidth,angle=0]{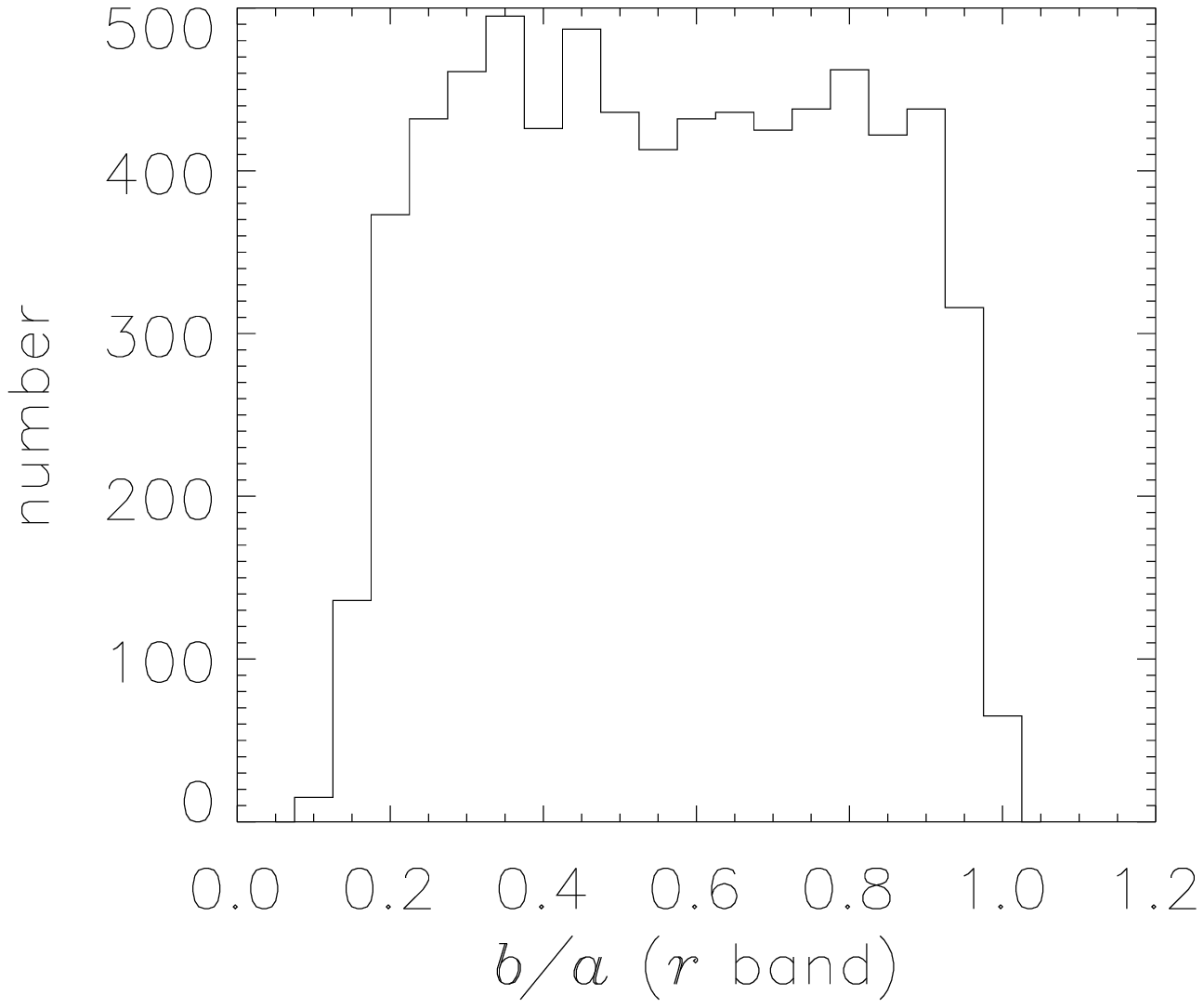}}
     \hspace{-1.5in}
     \subfigure[\redshift \ = \samplethreeredshift.]{
          \label{fig:z3}
          \includegraphics[width=.35\textwidth,angle=0]{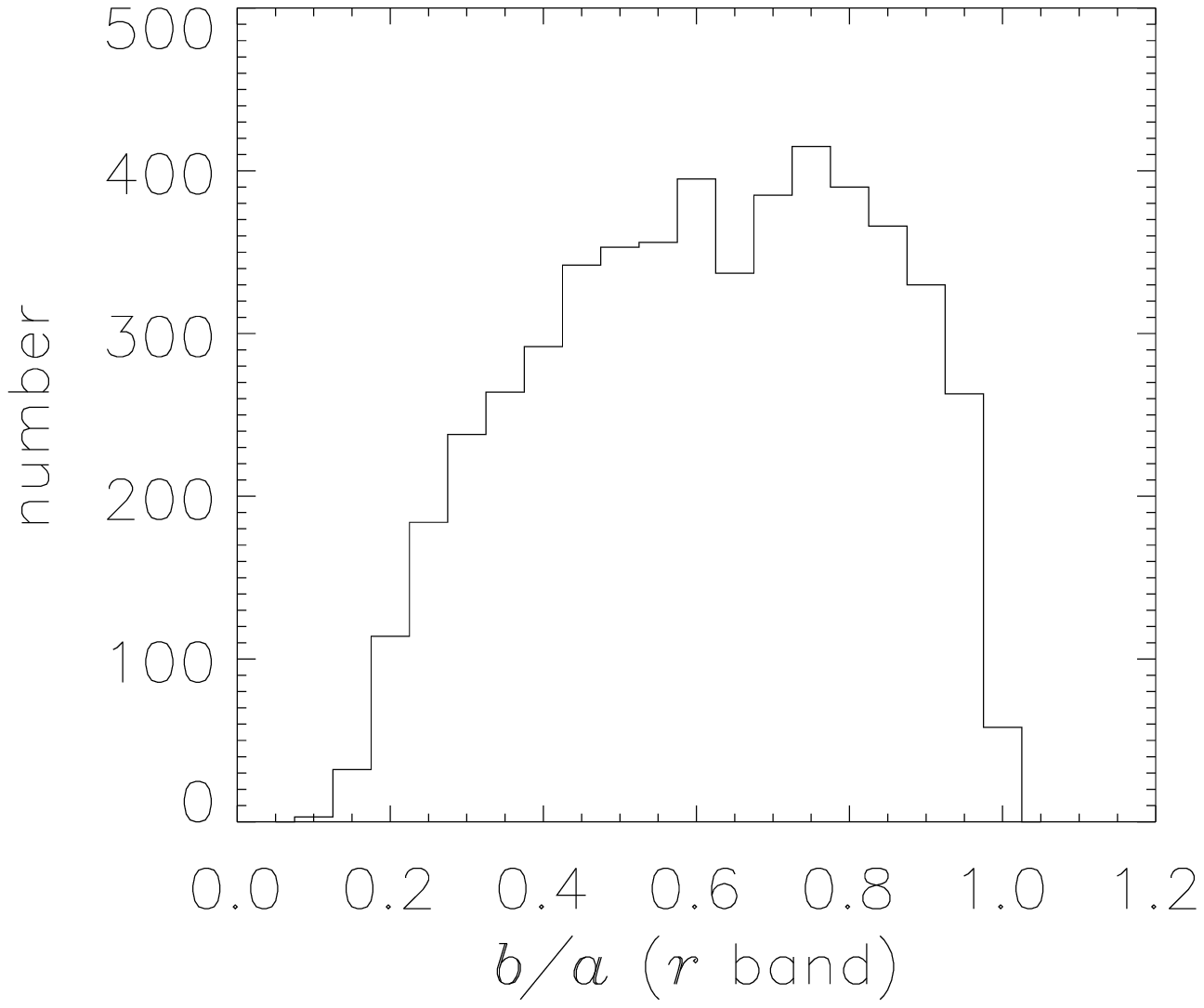}}
\end{subfigmatrix}
\caption{The number distribution of the axis ratio (\axisratio) of the
galaxies  in   our  parent   sample  in  the   range  \fracdevr   \  =
\samplefracdevr, at three different  slices of redshift. The narrowest
in  the   redshifts  is  designed   for  the  purpose   of  generating
volume-limited samples. Without additional magnitude cuts, we see that
at  lower redshifts  (\figname\ref{fig:z1}) the  face-on  galaxies are
relatively    under-sampled.    Conversely,   at    higher   redshifts
(\figname\ref{fig:z3})    the   edge-on   galaxies    are   relatively
under-sampled.}
\end{figure}

\begin{figure}
\epsscale{0.8}\plotone{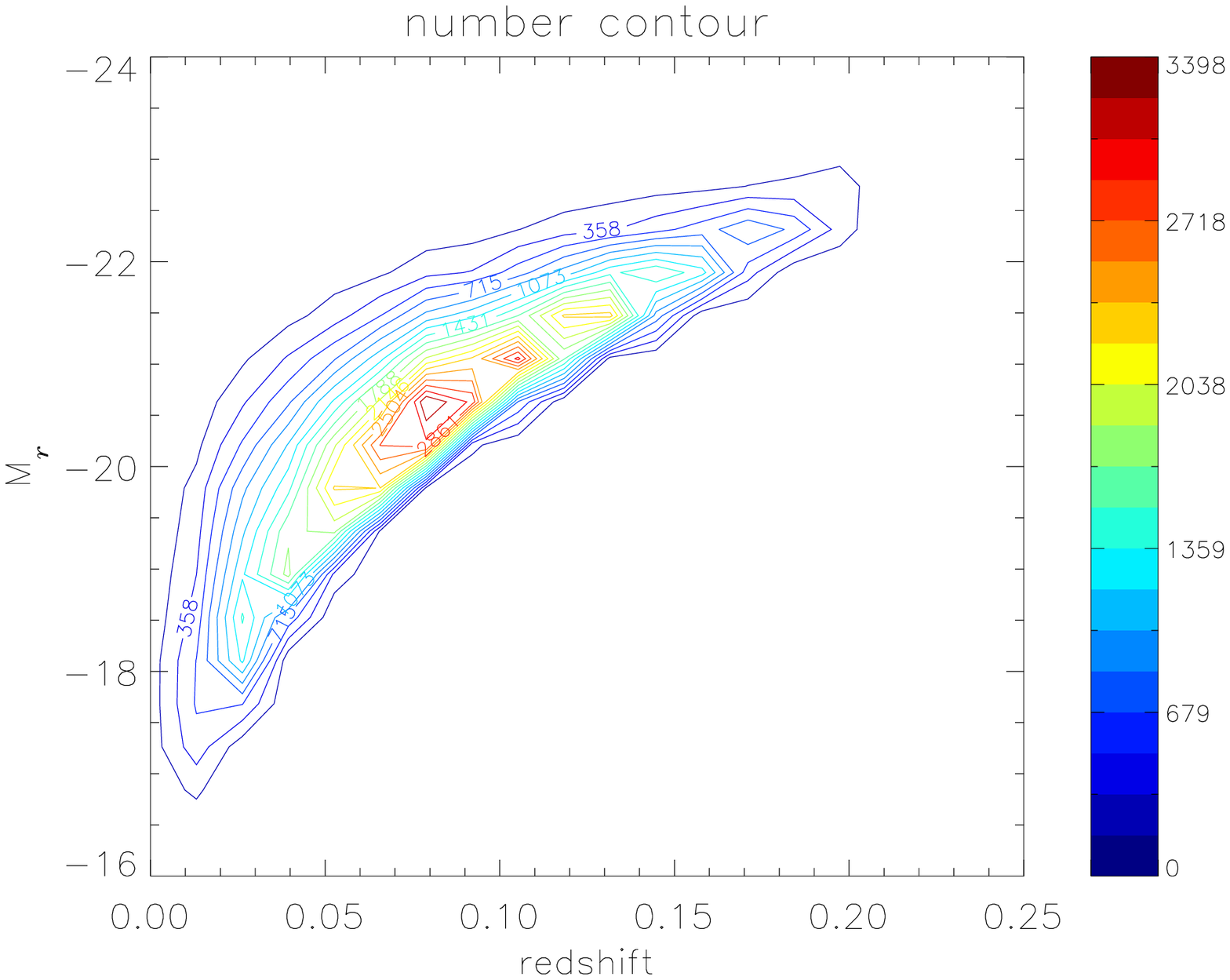}
\caption{The restframe $r$-band  Petrosian absolute magnitude vs.  the
redshift  of the  galaxies in  the  parent sample,  where \fracdevr  \
ranges  \ from  \samplefracdevr.  The  lower absolute  magnitude limit
corresponds to an apparent magnitude of \sdssmrlimit.}
\label{fig:psample1_Mr_sdss_redshift_fracdevr_cut}
\end{figure}

To continue with the sample  selection we plot the number distribution
of the axis ratio of the galaxies (\axisratio) in \figname\ref{fig:z1}
to \ref{fig:z3}.  Solely based on redshift cuts and without additional
magnitude cuts on  top of the survey magnitude limits,  we see that at
lower     redshifts    (\figname\ref{fig:z1},     \redshift     \    =
\sampleoneredshift) the face-on galaxies are relatively under-sampled.
Conversely, at  higher redshifts (\figname\ref{fig:z3},  \redshift \ =
\samplethreeredshift)    the   edge-on    galaxies    are   relatively
under-sampled.   The upper  and  lower absolute  magnitude limits  are
illustrated                                                          in
\figname\ref{fig:psample1_Mr_sdss_redshift_fracdevr_cut},     as     a
function   of  redshift.    As  the   SDSS  spectroscopic   sample  is
flux-limited (\sdssr-band  Petrosian apparent magnitude  brighter than
\sdssmrlimit), we deduce  that the lack of edge-on  galaxies at higher
redshifts is due to the  dropping out of the survey apparent magnitude
limit for those  highly inclined objects. At the  redshift of 0.1, the
highly  inclined objects  that  are dropped  out  of the  flux-limited
survey         would        be,        as         inferred        from
\figname\ref{fig:psample1_Mr_sdss_redshift_fracdevr_cut},
intrinsically  dimmer than $\approx~${-20}.   Conversely, the  lack of
face-on galaxies at  lower redshifts is expected to  be related to the
interplay  among all  of  the selection  biases described  previously,
causing more edge-on systems to be picked.

\begin{figure}
\epsscale{0.8}\plotone{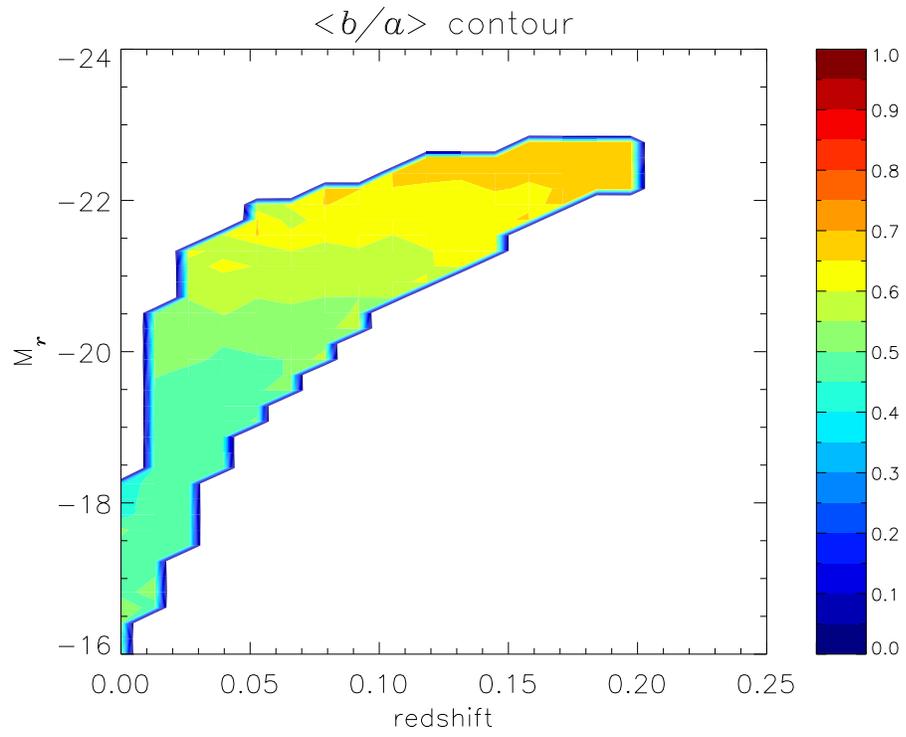}
\caption{The  average  inclination  of  the galaxies  (parent  sample,
\fracdevr \ = \samplefracdevr), plotted in the plane of \Mrpetro \ vs.
redshift.    The  corresponding  number   distribution  is   shown  in
\figname\ref{fig:psample1_Mr_sdss_redshift_fracdevr_cut}.  }
\label{fig:psample1_Mr_sdss_redshift_expab_fracdevr_cut}
\end{figure}

To  understand   further  this   result,  we  calculate   the  average
inclination of  the galaxies, plotted in  the plane of  \Mrpetro \ vs.
redshift
(\figname\ref{fig:psample1_Mr_sdss_redshift_expab_fracdevr_cut}).
Only galaxies  at $z  \approx $ \sampletworedshift  \ show  an average
inclination  of $\approx$~0.5,  consistent with  the  expected average
inclination of  a random sample  of galaxies that is  unbiased against
inclination and  the range  of the inclination  being within 0  and 1.
This consistency,  we note, does  not {\it prove} that  those galaxies
are  free of  any selection  bias.  One  such indication  is  that the
average  \ba  \  value  for  brighter  galaxies  (brighter  than  M$_r
\approx~$-21)  is not  0.5, but  0.65.  This  bias, however,  does not
affect  the  average luminosity  in  each  of  the composite  spectra,
because  the  average  luminosity  is  heavily  weighted  towards  the
characteristic luminosity of a galaxy sample and less sensitive to the
number  of  the  bright-end  objects.   Besides,  our  lower  absolute
magnitude limit,  -19.5, is dimmer than  the characteristic magnitudes
(the median \Mrpetro  \ ranged from -20.1 to -20.4 for  \ba \ from 0.1
to   1.0  for   galaxies   at   $z  =   $   \sampletworedshift  \   in
\figname\ref{fig:psample1_Mr_sdss_redshift_expab_fracdevr_cut}),
meaning  that  the  characteristic   luminosity  is  included  in  the
luminosity distribution.

 Because  the $z =  $ \sampletworedshift  \ sample  is mostly  flat in
terms  of  its  inclination distribution  (\figname\ref{fig:z2}),  the
relatively  small number  of nearly  edge-on ($b/a  = 0  - 0.1  $) and
nearly face-on ($b/a  = 0.9 - 1$) galaxies  in \figname\ref{fig:z2} is
interpreted   to   be    physical.    Discussed   by   other   authors
\citep[e.g.,][]{2007ApJ...659.1159S,2008ApJ...687..976U},     in    an
unbiased sample the lack of  exactly or nearly face-on galaxies is due
to intrinsic ellipticity of the galaxies \citep[on average $\epsilon =
1 - \ba =0.16$ in the SDSS disk galaxies,][]{2004ApJ...601..214R}, and
the lack  of exactly  or nearly edge-on  galaxies is due  to intrinsic
thickness of the galaxy disks.  Another factor that may come into play
is that the axis ratio \ba \ may not be a pure inclination measurement
of the  disk component  only, but of  the combined  bulge+disk systems
which are subjected  to the seeing effect. This  factor could explain,
e.g.,  the lack  of edge-on  galaxies.  In  the following,  we confine
ourselves to the galaxies  located at redshift \sampleredshift, with a
corresponding     survey      absolute     magnitude     limits     of
$\approx$~\samplemagnitude \ in the \sdssr \ band.

Finally,  only spectroscopically-classified star-forming  galaxies are
considered, where narrow-line active  galactic nuclei are removed from
the  sample   through  the  line  ratios   [\ion{O}{3}]/\Hbeta  \  vs.
[\ion{N}{2}]/\Halpha    \   according    to    the   calibration    by
\citet{2006MNRAS.372..961K}.  In  doing so the origin  of the emission
lines  of the  galaxies considered  in  the later  session, e.g.,  the
Balmer decrement  (\balmerdec), is well constrained --  arise from the
\ion{H}{2}  region that  is excited  by the  O/B stars.   We  take the
equivalent widths (EW's) of these  emission lines as those measured by
the  SDSS,  and  add   a  constant  restframe  stellar  absorption  of
\stellarabs     \     to      each     of     the     Balmer     lines
\citep{2003ApJ...599..971H,2003ApJ...597..142M}.  Each of the involved
emission  lines is  required to  be larger  than \restEWMin  \  in the
restframe  EW, and  is of  at  least \nsigmaline  \ detection.   These
selection criteria  provide higher confidence to  the Balmer decrement
measurements          that          are          performed          in
\S\ref{section:balmerdec}.   Composite  spectra  are   constructed  in
several  bins  of  inclination:  $\ba  =  0.1-0.2,  0.2-0.3,  0.3-0.4,
0.4-0.5, 0.5-0.6, 0.6-0.7, 0.7-0.8,  0.8-0.9,$ and $0.9-1.0$.  The bin
\badrop \ is dropped in the subsequent analysis because of the limited
number of galaxies (\tabname~\ref{tab:number}).

\subsection{Fraction of Galaxy Light through Spectroscopic Fiber} \label{section:fiber}

\begin{figure}\begin{center}
\epsscale{0.8}\plotone{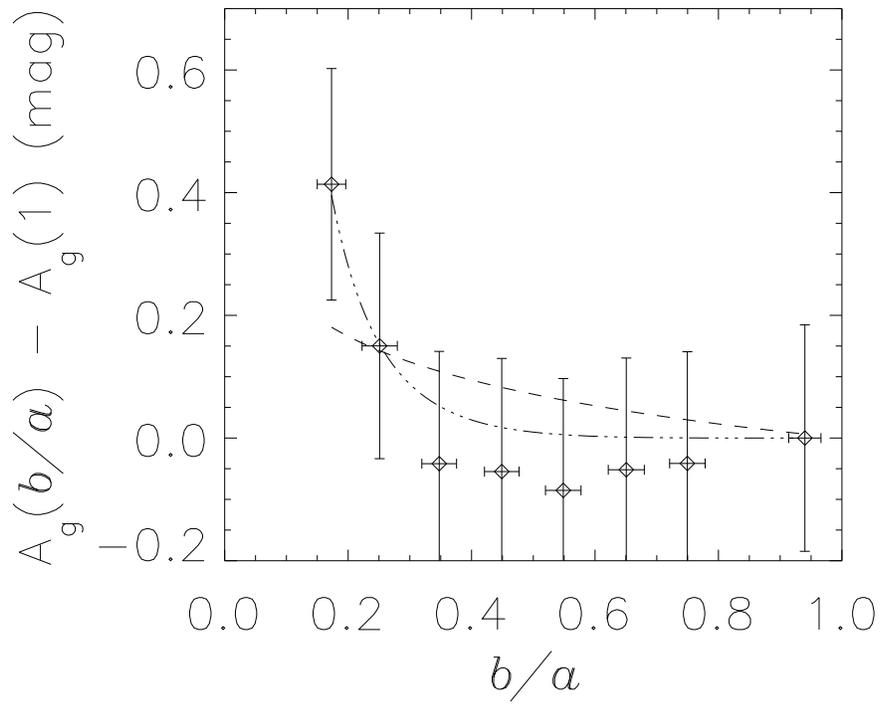}
\caption{Dependency of  the $g$-band  relative extinction in  the disk
  galaxies on  their inclination, for the  volume-limited sample.  The
  best-fit empirical models  are the log model (dashed  line), and the
  log$^4$ model (dot-dot-dot-dashed line).}
\label{fig:ext_inclin_g.bayesian.empiricalmodel}
\end{center}\end{figure}

The apparent  half-light radius of  the star-forming disk  galaxies is
found  to   increase  with  the   galaxy  inclination,  as   shown  in
\figname\ref{fig:psample1_expradr_vs_expabr_vlim_Mr_sdss_fracdevr_cut}.
Similar to  \citet{2009ApJ...691..394M}, the relation  in our galaxies
can be described by

\begin{equation}
{\log}_{10}  \left(   {\reffm}^{\ba}  \right)  =   {\log}_{10}  \left(
{\reffm}^{1} \right) - \beta_{r} \, {\log}_{10} \left( \ba \right) \ ,
\label{eqn:reff}
\end{equation}

\noindent
where  ${\reffm}^{\ba}$  is  the  apparent half-light  radius  of  the
inclined  galaxies, and  ${\reffm}^{1}$ is  the value  of  the face-on
galaxies. Through a linear  least-square fit, the best-fit coefficient
in the \sdssr \ band, $\beta_r$, is found to be $\betar \pm \betarsd$.
This     value     agrees    fairly     well     with    that     from
\citet{2009ApJ...691..394M}, who found $\beta_r = 0.2$ for a different
sample  of nearby  SDSS disk  galaxies which  also  showed inclination
dependency  in   their  observed  properties.    The  larger  apparent
half-light radius in the more  inclined disk galaxies was seen also in
\citet{1992MNRAS.254..677H}      and      \citet{2006A&A...456..941M},
respectively in an independent galaxy sample and in a model for galaxy
spectra.  The  adoption of  the relation in  \eqnname\ref{eqn:reff} is
physically  motivated.   Its  origin  is  presumably  related  to  the
extinction radial  gradient in the  galaxies, so that in  the inclined
galaxies the  central region dims by  a greater degree  than the outer
regions, resulting  in an increased apparent  half-light radius.  This
explanation   is   consistent   with   what  we   found   later   (see
\figname\ref{fig:ext_inclin_g.bayesian.empiricalmodel}),  that dust is
present  at least  in the  disk  galaxies, and  more so  for the  more
inclined galaxies.

Correcting   the   half-light  radius   of   each   galaxy  by   using
\eqnname\ref{eqn:reff}, the sample-average ratio between the radius of
the SDSS spectroscopic fiber (a  constant \rfiber $ = 1.5$\arcsec) and
the face-on  \reff \ (i.e.,  \rfiber/\reff$^{1}$) is calculated  to be
approximately   \rfibervsreffintrinsic.   This   ratio   stays  fairly
constant with the galaxy inclination, implying that a similar fraction
of  the  full galaxy  light  is measured  in  the  spectra of  various
inclinations.

\section{Inclination Dependency of Disk Composite Spectrum} \label{section:inclination}

 A composite spectrum  of the disk galaxies is  constructed in each of
the  bins of inclination.   The flux  density ($f_{\lambda}$)  of each
disk   galaxy   spectrum    is   converted   to   luminosity   density
($L_{\lambda}$)  by  multiplying  with  the  standard  factor  $4  \pi
d_{L}^2$, where $d_{L}$  is the luminosity distance of  the galaxy, so
that  a  fair  comparison  among  the  composites  can  be  undertaken
regardless of the redshift of  the galaxies.  This importance was also
pointed  out in \citet{1991Natur.353..515B}.   We use  the concordance
cosmological   parameters\footnote{The  cosmological   parameters  are
$\Omega_{V}  =  0.73,  \Omega_{M}  =  0.27$,  and  $h  =  0.71$.}   in
calculating $d_{L}$.  The flux density of a composite spectrum in each
wavelength bin  is taken to  be the geometric  mean of those  from the
contributing spectra, which preserves  the average of the steepness of
the spectral slopes  in a sample \citep[e.g.,][]{2001AJ....122..549V},
and  is  effective  in   removing  noisy  data.   Following  the  SDSS
convention, the composite spectra are expressed in vacuum wavelengths,
and  are  rebinned  into   $\waveS  -\waveE$~\AA  \  with  a  spectral
resolution of 70~\kms \ per wavelength bin.

The   comparison   between    the   composite   spectra   of   face-on
$(\ba=0.9-1.0)$ and edge-on $(\ba=0.1-0.2)$  disk galaxies is shown in
\figname\ref{fig:flux_edgeon_faceon_0.1_0.2.gmean.3700._5000.}      and
\ref{fig:flux_edgeon_faceon_0.1_0.2.gmean.5000._8000.}.     A   marked
difference  is  seen in  the  luminosity  and  the slope  between  the
continua, as such the inclined  galaxies show a lower luminosity and a
smaller slope in  the continuum.  The effect is  in agreement with the
usual   power-law   wavelength   dependence   in   extinction   curves
\citep[e.g.,][]{1994ApJ...429..582C}.

\section{Continuum Extinction in \lowercase{$g,r,i$} bands} \label{section:extinction}

\subsection{Relative Continuum Extinction}

Assuming the inclination dependency in the composite spectra is mainly
due to  extinction in  the samples of  the galaxies, we  calculate the
extinction   of  the   inclined   disk  at   the   filter  band   $x$,
$A_{x}(\ba)$~(mag), by comparing  the inclined with face-on luminosity
($L_{x}(\ba)$  and  $L_{x}(1)$)  derived  from the  composite  spectra
according to

\begin{equation}
A_{x}(\ba)  -  A_{x}(1) =  -2.5  \log_{10} \left[  L_{x}(\ba)/L_{x}(1)
  \right] \ , \label{eqn:relext}
\end{equation}

\noindent 
where  $A_{x}(1)$  is   the  face-on  extinction.   The  corresponding
relative optical  depth ($\tauxinclin - \tauxfaceon$)  can be obtained
by  dividing the extinction  with the  factor $2.5\,  \log_{10}(e)$ or
$1.086$. Although one can get only the inclined extinction relative to
the face-on  value but  not both individually,  the advantage  of this
approach is its  being empirical. Throughout the paper,  we denote the
face-on inclinations to be $\ba = 1$ for simplicity, where the highest
inclinations   are   in   fact   $0.94\pm   0.02$   for   our   sample
(Table~\ref{tab:extrel}).

The  broad-band  luminosity  is  derived  by  convolving  a  composite
spectrum  with  a  given  filter  response curve.   To  focus  on  the
extinction in the stellar populations instead of other components such
as  the \ion{H}{2} regions,  the flux  contribution from  any emission
line  to the  synthetic magnitudes  are  excluded.  To  make sure  the
continuum extinction is not biased by the continuum estimation method,
we  use  and compare  two  independent  approaches:  (1) an  empirical
sliding window  approach (2)  a Bayesian approach  based on  a stellar
population model.

In the sliding window approach,  the continuum luminosity density at a
given restframe  wavelength is estimated to be  the average luminosity
density  between   the  40th  and  60th  percentiles   of  the  sample
distribution of  the contributing  luminosity densities, which  are in
turn  taken from  within $\pm  150$ wavelength  bins (in  the spectral
resolution of 70~\kms  \ per bin) around the  centered wavelength. The
emission lines are  masked out within a constant  wavelength window of
$\pm   280$~\kms,   utilizing   the   line   list   in   Table~30   of
\citet{2002AJ....123..485S}.

In    the    Bayesian     approach,    the    method    outlined    in
\citet{2007MNRAS.381L..74K}     is     adopted.      We    use     the
\citet{2003MNRAS.344.1000B} high-resolution  model (1~\AA \ resolution
in the restframe wavelengths considered in this work) to determine the
integrated  stellar   light  of  each  composite   spectrum,  and  the
extinction   model  by   \citet{2000ApJ...533..682C}   for  the   dust
reddening.   The  star  formation   history  is  characterized  by  an
exponential  declining star  formation  rate defined  by an  e-folding
time,  and the  model parameters  are, the  age of  the  oldest stars:
$1-13.7$~Gyr,  the  stellar   metallicity  ($Z$):  $0.0004-0.05$,  the
e-folding  time:   $1-15$~Gyr,  and  the   nebular  gas  emission-line
reddening: $0-0.8$~mag.

For  space  saving,  not  all  of the  results  using  both  continuum
estimation methods are presented. However,  we stress that there is no
substantial  difference detected  in the  continuum  extinction values
between both approach.  For unity,  we present and discuss our results
in  this  work based  on  the  Bayesian  continuum estimation,  unless
otherwise  specified.   One  significant  advantage  of  the  Bayesian
approach  is in  the Balmer-line  measurements,  as will  be shown  in
\S\ref{section:balmerdec},  where  the   flux  contribution  from  the
underlying  stellar absorption is  represented automatically  by using
the best-fit stellar population model spectrum.

In \figname\ref{fig:ext_inclin_g.bayesian.empiricalmodel}  we see that
the  relative $g$-band  extinction $A_{g}(\ba)  -  A_{g}(1)$ increases
with the  inclination of the  galaxies.  The relative  extinctions for
the volume-limited sample are given in Table~\ref{tab:extrel}.  In the
highly inclined objects ($\ba  = 0.26\pm0.03$) $A_{g}(\ba) - A_{g}(1)$
is   \rextginclin,  $A_{r}(\ba)   -   A_{r}(1)$  is \rextrinclin,   and
$A_{i}(\ba) - A_{i}(1)$ is \rextiinclin.

\subsection{Empirical Extinction Model} \label{section:empiricalmodel}

To  extrapolate the  relative  extinction to  even higher  inclination
values where no  result is available, empirical models  are used.  The
$\log(\ba)$                                                       model
\citep{1991trcb.book.....D,1995A&A...296...64B,1998AJ....115.2264T} is

\begin{equation}
A_{x}(\ba) - A_{x}(1) = - \gamma_{x} \, \log_{10} \left( \ba \right) \ .
\end{equation}

\noindent
The         best-fit         model         is         drawn         in
\figname\ref{fig:ext_inclin_g.bayesian.empiricalmodel}.  As found also
by \citet{2008ApJ...687..976U}, the actual dependence is found here to
be steeper than  that described by the $\log(\ba)$  model. Instead, we
found the following

\begin{equation}
A_{x}(\ba) - A_{x}(1) = \eta_{x} \, \log_{10}^4 \left( \ba \right)
\end{equation}

\noindent
 provides a better fit to our result. This choice is purely empirical,
as a  physically motivated  model depends on  knowledge of  the actual
distribution of stars and dust in  a galaxy. In the \sdssg \ band, the
best-fit $\eta_{g} = $\rextg.   This value gives a relative extinction
of \rextg \ for highly  inclined galaxies with $\ba = \baedgeonextra$,
agrees      with     those      from     the      previous     studies
\citep{2007ApJ...659.1159S,2009ApJ...691..394M}     at     the    same
inclination, both being $\gsim{1}$~mag in the \sdssg \ band.

\subsection{Theoretical Extinction Model: Face-On Extinction} \label{section:theoreticalmodel}

To    derive   the    face-on    ($\ba   =    1$)   extinction,    the
inclination-dependent continuum extinction  is fitted with theoretical
models  which  describe  the   distributions  of  stars  and  dust  in
individual galaxies. The screen model is

\begin{equation}
A_{x}(\ba) = A_{x}(1) / (\ba) \ .
\label{eqn:screenmodel} 
\end{equation}

\noindent
In this scenario, the absorbing layer of material is located above the
stars  like  a  screen. This  choice  is  motivated  by the  study  of
\citet{2007ApJ...659.1159S} where  the authors found  that the optical
depth of  inclined disk  galaxies in the  SDSS is proportional  to the
cosine of the inclination angle, which is in accord to the description
by the screen model.  Next,  the slab  model is
considered

\begin{equation}
A_{x}(\ba) = -2.5 \, {\log}_{10} { \left[ {\frac{\ba}{\tauxfaceon}} \,
  \left( 1 - e^{- \tauxfaceon/(\ba)} \right) \right] } \ ,
 \label{eqn:slabmodel} 
\end{equation}

\noindent
for which the  stars and the dust are mixed uniformly  in a slab. Finally,
the sandwich model is

\begin{eqnarray}
A_{x}(\ba)  & =  & -2.5  \,  {\log}_{10} {\left[  \frac{1 -  \zeta}{2}
    \left( 1  + e^{-\tauxfaceon/(\ba)}  \right) \right.  }   \\ &  & +
    \left.    \frac{\zeta   *   (   \ba)}{\tauxfaceon}  \left(   1   -
    e^{-\tauxfaceon/(\ba)} \right) \right] \ ,
 \label{eqn:sandwichmodel} 
\end{eqnarray}

\noindent
for which a  layer of dust+stars mixture is  sandwiched in-between two
layers of stars.   In these models the face-on  extinction and the optical
depth are respectively $A_{x}(1)$  and $\tau_{x}(1)$.  The geometry of
these models are illustrated in detail in \citet{1989MNRAS.239..939D}.
Our  model choice for  dust+stars is  non-exhaustive, there  are other
models in the literature which  address, e.g., the presence of clumped
dust \citep[e.g.,][]{2004A&A...419..821T}, which we plan to explore in
the future.

\begin{figure}\begin{center}
\epsscale{0.8}\plotone{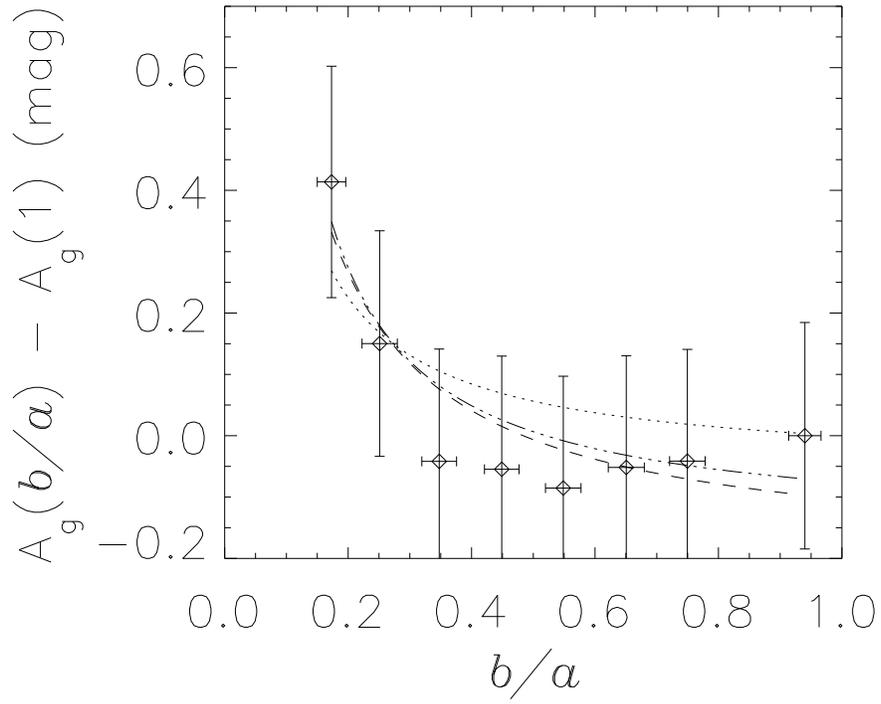}
\caption{Dependency of  the $g$-band  relative extinction in  the disk
  galaxies on  their inclination, for the  volume-limited sample.  The
  best-fit theoretical models are  the screen model (dotted line), the
  slab model (dashed line)  and the sandwich model (dot-dot-dot-dashed
  line).}
\label{fig:ext_inclin_g.bayesian}
\end{center}\end{figure}

The         best-fit        models         are         drawn        in
\figname\ref{fig:ext_inclin_g.bayesian}.   The  corresponding best-fit
face-on  extinctions   in  the  SDSS   $g,r,i$  bands  are   given  in
Table~\ref{tab:extfaceon}. Among all  models the screen model performs
the least  satisfactory, whereas the slab and  sandwich models perform
similarly.  The best-fit values  of $A_{g}(1)$ are model-dependent, as
pointed  out   also  by  \citet{1989MNRAS.239..939D}.    The  best-fit
$A_{g}(1)$ however are  similar in both the slab  and sandwich models,
$\sim{0.2}$~mag.

\subsection{Extinction Curves in the Optical} \label{section:extcurve}

The wavelength  dependence of the  relative extinction is  obtained by
comparing the inclined to face-on composite spectra.  In the following
discussions,  the inclined  spectrum is  chosen  to be  that from  the
highest   inclinations   ($\ba   =  \baedgeonrange$),   the   detailed
inclination dependence of the relative extinction in the continuum can
be obtained,  e.g., by invoking the best-fit  empirical or theoretical
models                                 (\S\ref{section:empiricalmodel},
\S\ref{section:theoreticalmodel}).   Within  the restframe  wavelength
range $\waveS  -\waveE$~\AA, the relative extinction as  a function of
wavelength is calculated  according to \eqnname\ref{eqn:relext}, where
the broad-band luminosity on  the left-hand-side is now substituted by
the  luminosity  density  ($L_{\lambda}$)  of the  composite  spectrum
multiplied by  the size of  the wavelength bin  ($\Delta{\lambda}$) at
the concerned wavelength.  The  resultant extinction curve is shown in
\figname\ref{fig:extin_0_0.3.ndeg.3.3700._5000.}    for   $\lambda   =
\waveS                -               \waveMid$~\AA,               and
\figname\ref{fig:extin_0_0.3.ndeg.3.5000._8000.}    for   $\lambda   =
\waveMid  - \waveE$~\AA.   Comparing  the relative  extinction in  the
lines  and  in the  stellar  continuum,  the  following cases  can  be
visually identified:

\begin{description}
\item[$\dagger$] lines  which exhibit relative  extinction larger than
that in  the surrounding stellar  continuum: Ca\,K and  H absorptions,
Na\,D$\lambda{5896}$  absorption, \ion{He}{1}$\lambda{3889}$, \Halpha,
\Hbeta  , \Hgamma ,  \Hdelta \  emissions, [\ion{N}{2}]$\lambda{6550}$
and $\lambda{6585}$ emissions
\item[$\dagger$] lines which  exhibit relative extinction smaller than
  that      in       the      surrounding      stellar      continuum:
  [\ion{O}{3}]$\lambda{4960}$           and           $\lambda{5008}$,
  [\ion{O}{2}]$\lambda{3727}$       and       $\lambda{3730}$      and
  [\ion{O}{1}]$\lambda{6302}$ emissions
\item[$\dagger$]  lines which exhibit  relative extinction  similar to
  that      in       the      surrounding      stellar      continuum:
  [\ion{S}{2}]$\lambda{6718}$ and $\lambda{6733}$ emissions
\item[$\dagger$] uncertain due to small line strength in the composite
spectra:  \Heta,  \Htheta  \  emissions,  [\ion{O}{3}]$\lambda{4364}$,
[\ion{O}{1}]$\lambda{6366}$ emissions
\end{description}

\noindent
These lines  are all  present in the  face-on composite  disk spectra,
 albeit in the  last case the line strengths  in the composite spectra
 are too small for identification.  It is intriguing to see that there
 is a variety  of behaviors, in terms of  the inclination-induced dust
 extinction among the emission/absorption lines, and their relation to
 the extinction in the stellar  continuum.  In particular, none of the
 oxygen forbidden emission lines shows larger relative extinction than
 in the  stellar continuum.  Besides,  the relative extinction  in the
 [\ion{O}{3}]$\lambda{5008}$  is smaller  than that  in the  \Hbeta, a
 little bit  surprising if they  are emitted from the  same \ion{H}{2}
 region.  These results are expected  to give us hints on the geometry
 of the dust  in relation to the regions where  these lines arise, and
 on  the extinction  mechanism (e.g.,  type of  dust) on  each  of the
 elements.  This analysis  however is beyond the scope  of this paper,
 and will be  performed in a future work.  In this  work we focus only
 on the  Balmer decrement, \balmerdec, commonly used  for deriving the
 dust reddening  in \ion{H}{2}  region within a  emission-line galaxy,
 later in \S\ref{section:balmerdec}.

The wavelength-dependent relative  extinction in the stellar continuum
is next fitted by a polynomial function of the 3rd degree

\begin{equation}
A_{\lambda * \Delta\lambda}(\ba)  - A_{\lambda * \Delta\lambda}(1)  =  \sum_{j =  0}^{3}  \, a_{j}  \,
{\wavenumber}^{j} \ , \label{eqn:extcurve}
\end{equation}

\noindent
where   the  wavenumber   $\wavenumber$  (i.e.,   inverse  wavelength,
$\wavenumber  =  1/\lambda$)  is   in  the  unit  of  inverse  micron,
$\micron^{-1}$.     The   best-fit    coefficients   are    given   in
\tabname\ref{tab:extcurve}.

Assuming  the stellar population  and dust  properties of  the edge-on
galaxies  are  not  specially  different  from those  of  the  face-on
galaxies,    the   combination    of    \eqnname\ref{eqn:relext}   and
\eqnname\ref{eqn:extcurve} can be used  to correct against the effects
of reddening and  extinction in the observed continuum  of an inclined
star-forming disk galaxy.

\section{Balmer Decrement} \label{section:balmerdec}

The reddening on the emissions from the \ion{H}{2} regions such as the
Balmer decrement  (the intensity ratio  between \Halpha \  and \Hbeta)
depends  on   the  detailed   geometry  of  the   dust  and   the  gas
\citep[e.g.,][]{1986A&A...155..297C}.  If the reddening induced by the
galaxy inclination  affects the emissions from  the \ion{H}{2} regions
of the  galaxy in a  similar manner as  the integrated light  from the
continuum-generating  stars,  the  measured  \balmerdec  \  ratio  can
in-principle  depend on  the  inclination of  a  galaxy. We  therefore
explore  the  relationship  between   the  Balmer  decrement  and  the
inclination-dependent continuum extinction.

The Balmer line ratio \balmerdec  \ provides a mean for characterizing
the   dust  reddening   in  ``normal''   \ion{H}{2}   regions  (case~B
recombination, $n_e  = 100$~cm$^{-3}$, $T_e =  10,000$~K), because the
line  ratio  is  fairly   in-sensitive  to  the  electron  temperature
\cite[see Table~4.4 of][]{1989agna.book.....O}. In measuring the EW of
a Balmer line, we fit to the continuum-subtracted composite spectrum a
single Gaussian function.  In measuring  the uncertainty in an EW, the
following  steps  are  performed.   Firstly, the  uncertainty  in  the
luminosity  densities of the  stellar continuum  in each  composite is
calculated                           by                          using
${\delta{L_{\lambda}^c}/{L_{\lambda}^c}}={\delta{L_{\lambda}}/{L_{\lambda}}}$,
where $\delta$ denotes the uncertainty of a quantity.  The uncertainty
in   each   Balmer   EW   is  calculated   to   be   $\sum_{\lambda_R}
{\delta{(L_{\lambda} -  L_{\lambda}^c)}/{L_{\lambda}^c}} \, d\lambda$,
where $\lambda_R$  is the region  of influence of the  emission lines,
taken  to  be  around the  line  center,  $\pm  280$~\kms \  and  $\pm
200$~\kms, respectively  for the emission lines \Halpha  \ and \Hbeta.
Finally, the  uncertainty in the  \balmerdec \ is propagated  by using
$(\Halpham/\Hbetam)      \,      \sqrt((\delta\Hbetam/\Hbetam)^2     +
(\delta\Halpham/\Halpham)^2)$.

\begin{figure}\begin{center}
\plotone{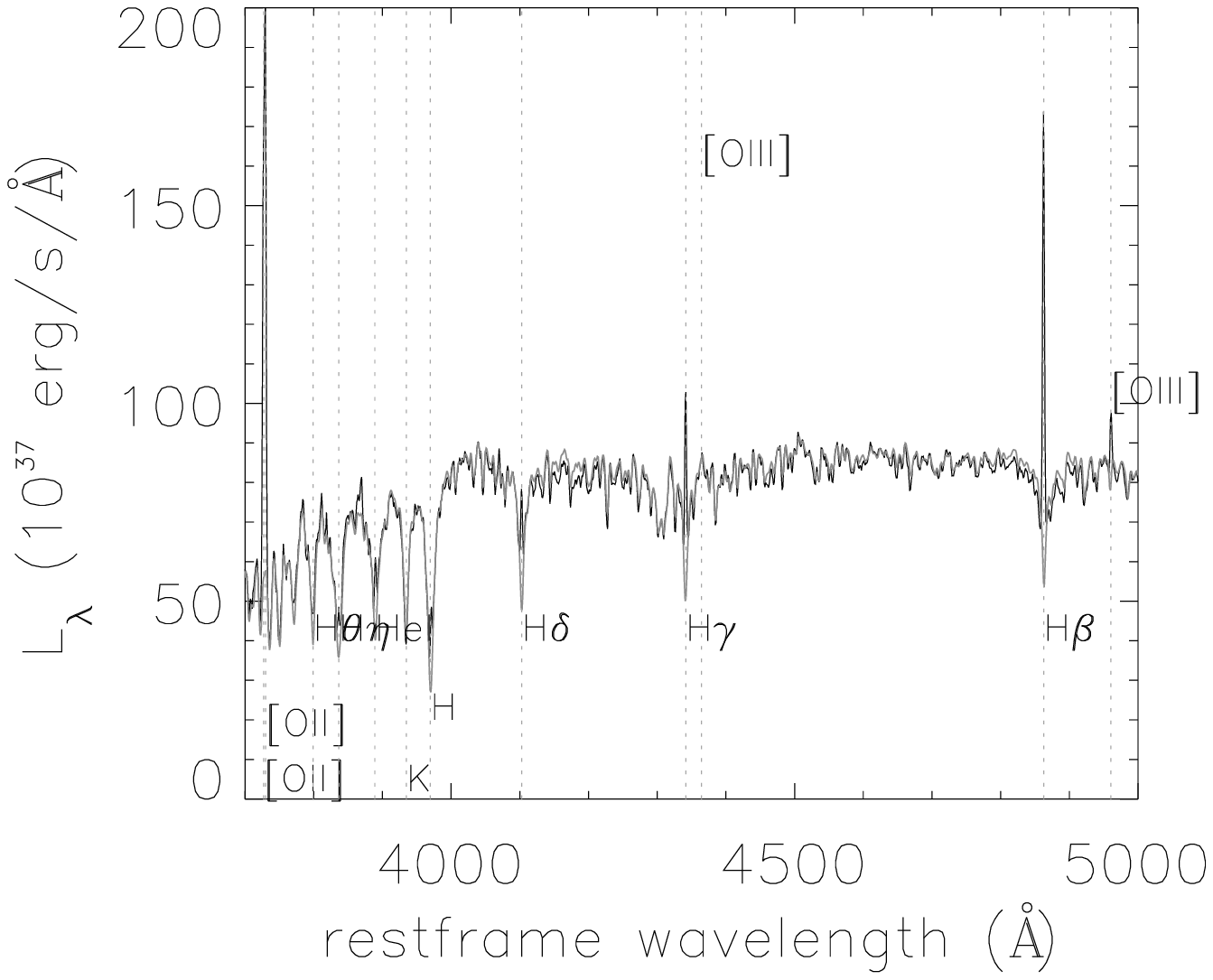}
\caption{The face-on ($\ba = \bafaceonrange$) composite spectrum
  (black) and the best-fit stellar continuum
using the Bayesian approach (gray), in the vicinity of the \Hbeta.}
\label{fig:linefit_hbeta}
\end{center}\end{figure}

\begin{figure}\begin{center}
\plotone{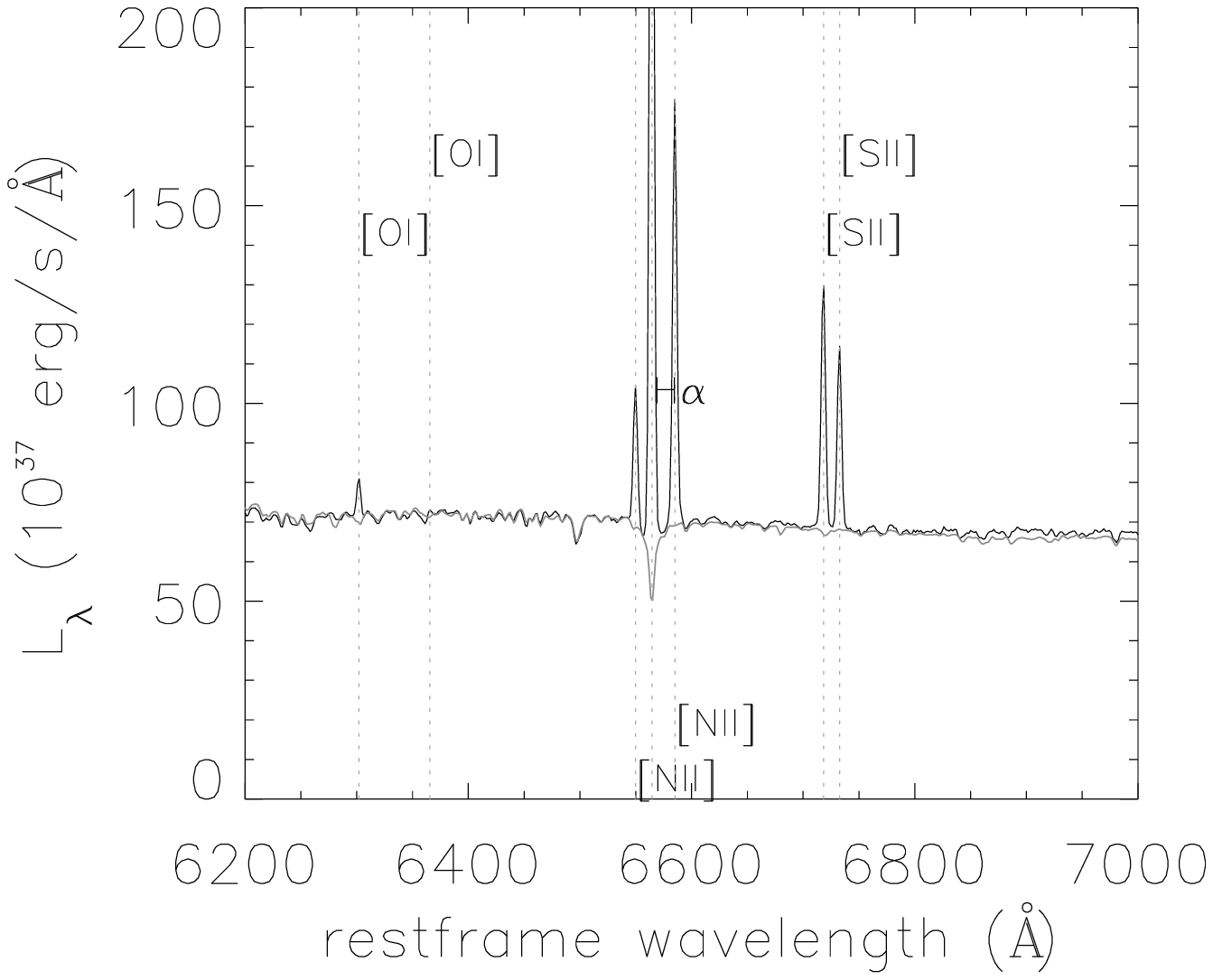}
\caption{The face-on ($\ba = \bafaceonrange$) composite spectrum
  (black) and the best-fit stellar continuum
using the Bayesian approach (gray), in the vicinity of the \Halpha.}
\label{fig:linefit_halpha}
\end{center}\end{figure}

Examples of the best-fit stellar  continua in the vicinity of \Hbeta \
and    \Halpha    \    emissions    are    shown    respectively    in
\figname\ref{fig:linefit_hbeta}  and \figname\ref{fig:linefit_halpha}.
Using  the  Bruzual  and  Charlot  stellar population  model  and  the
Calzetti dust  model, in the Bayesian approach  the underlying stellar
absorption lines are represented obviously in the best-fit continuum.

\begin{figure}\begin{center}
\plottwo{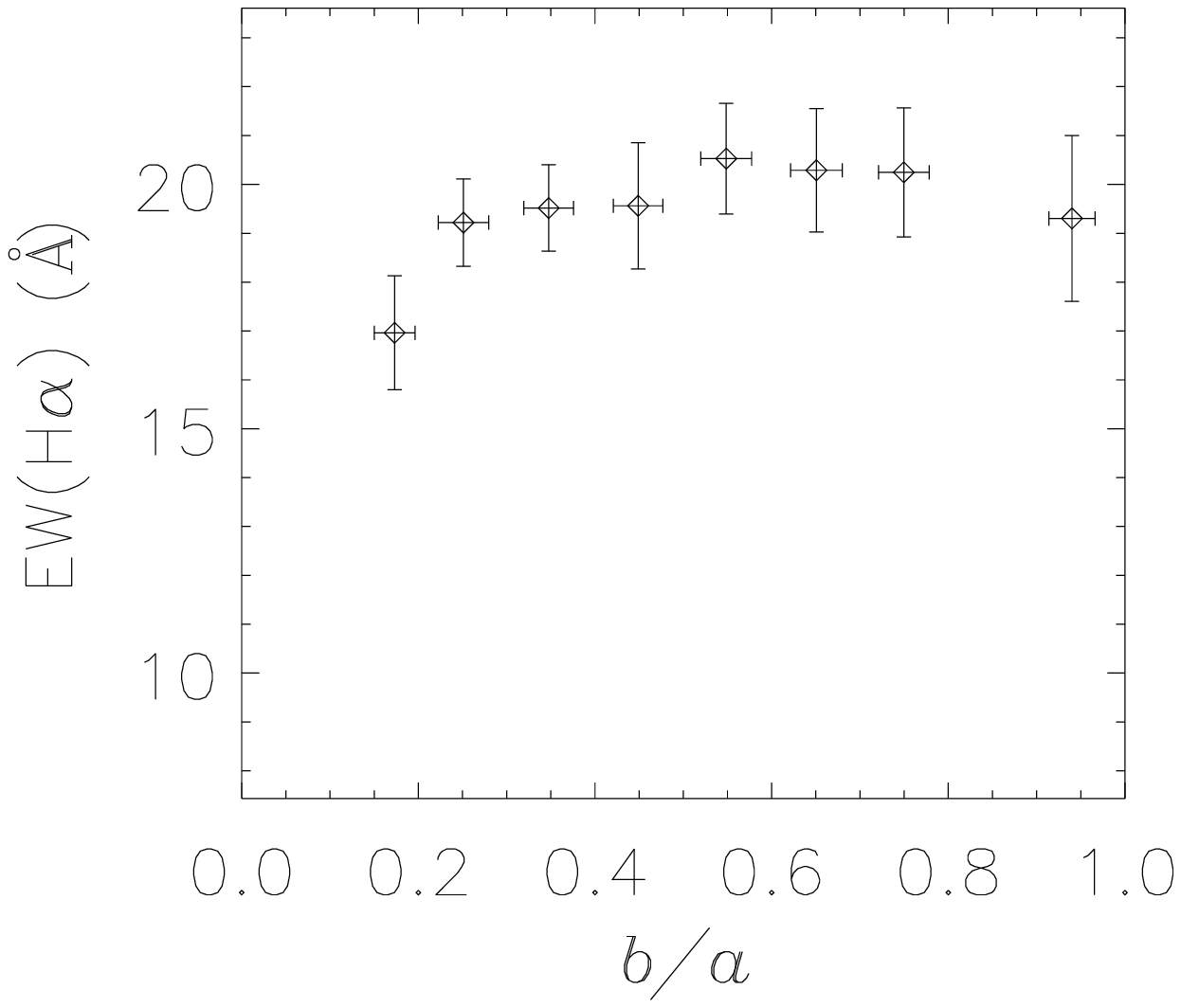}{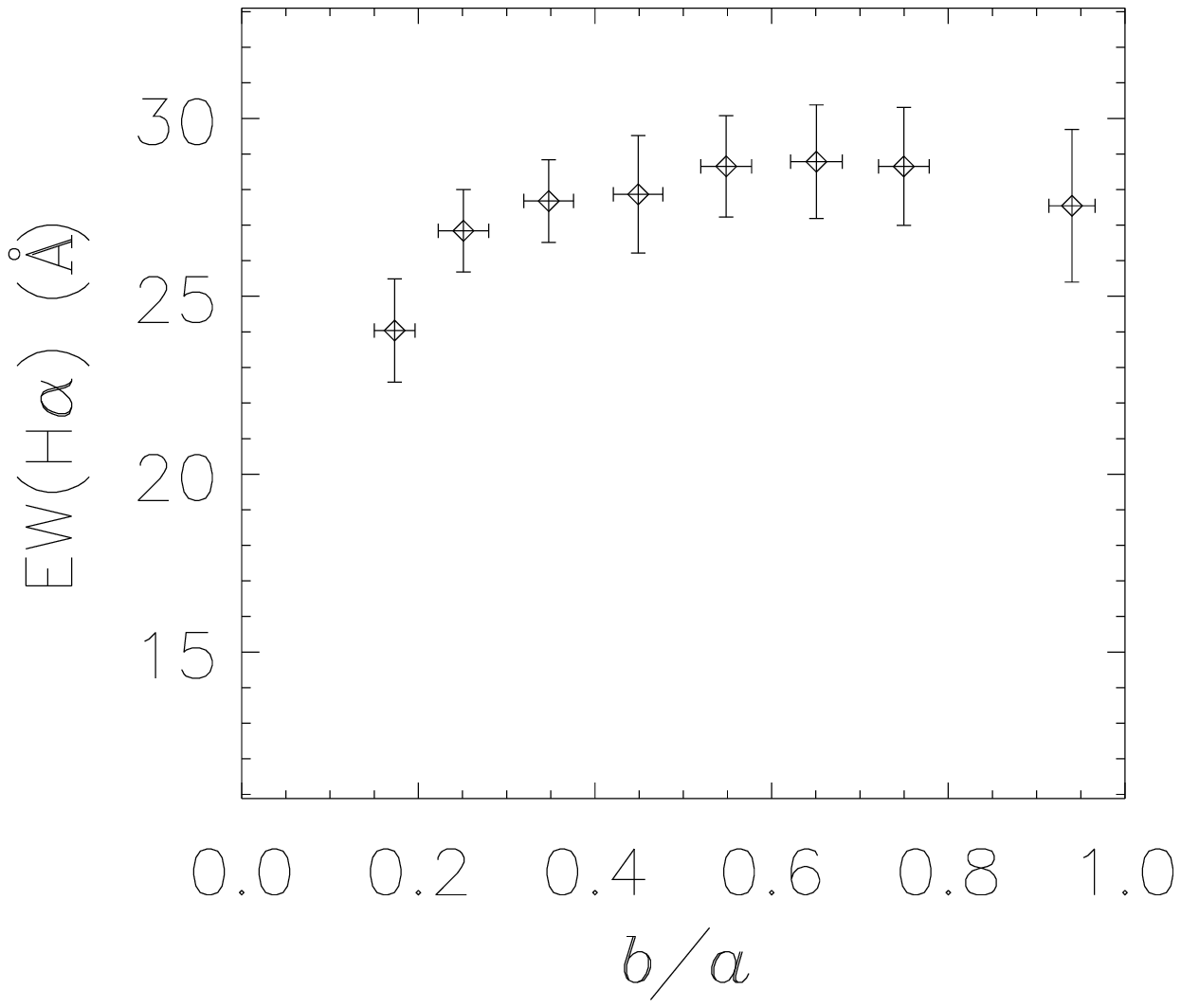}
\caption{The  inclination  dependency  of  the  \Halpha  \  equivalent
width. The  sliding window approach  (left) and the  Bayesian approach
(right) are  used respectively in estimating the  stellar continuum in
the composite spectra. The error bar is \onesigma \ statistical
uncertainty for each parameter. }
\label{fig:ha_ew_inclin}
\end{center}\end{figure}

\begin{figure}\begin{center}
\plottwo{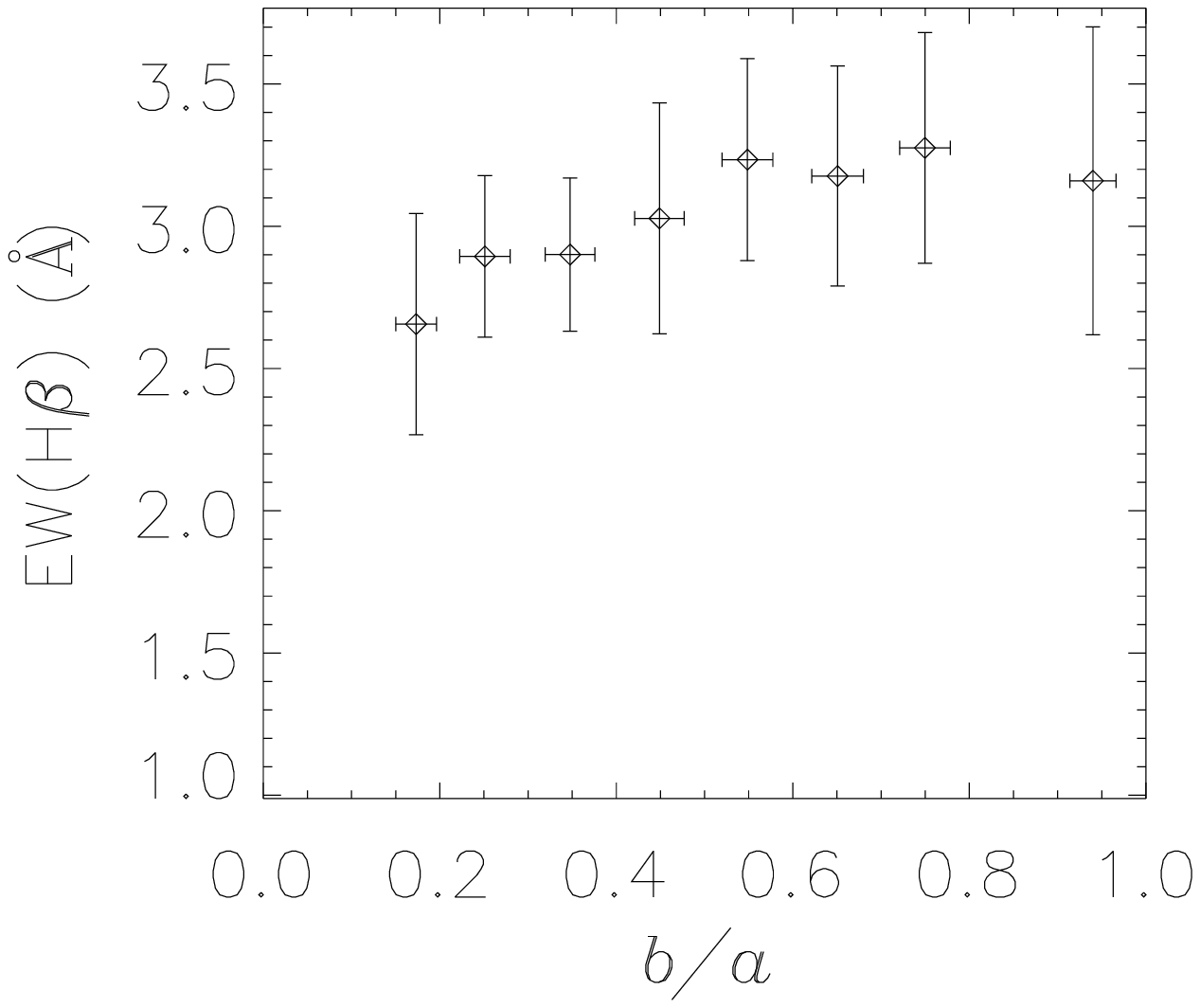}{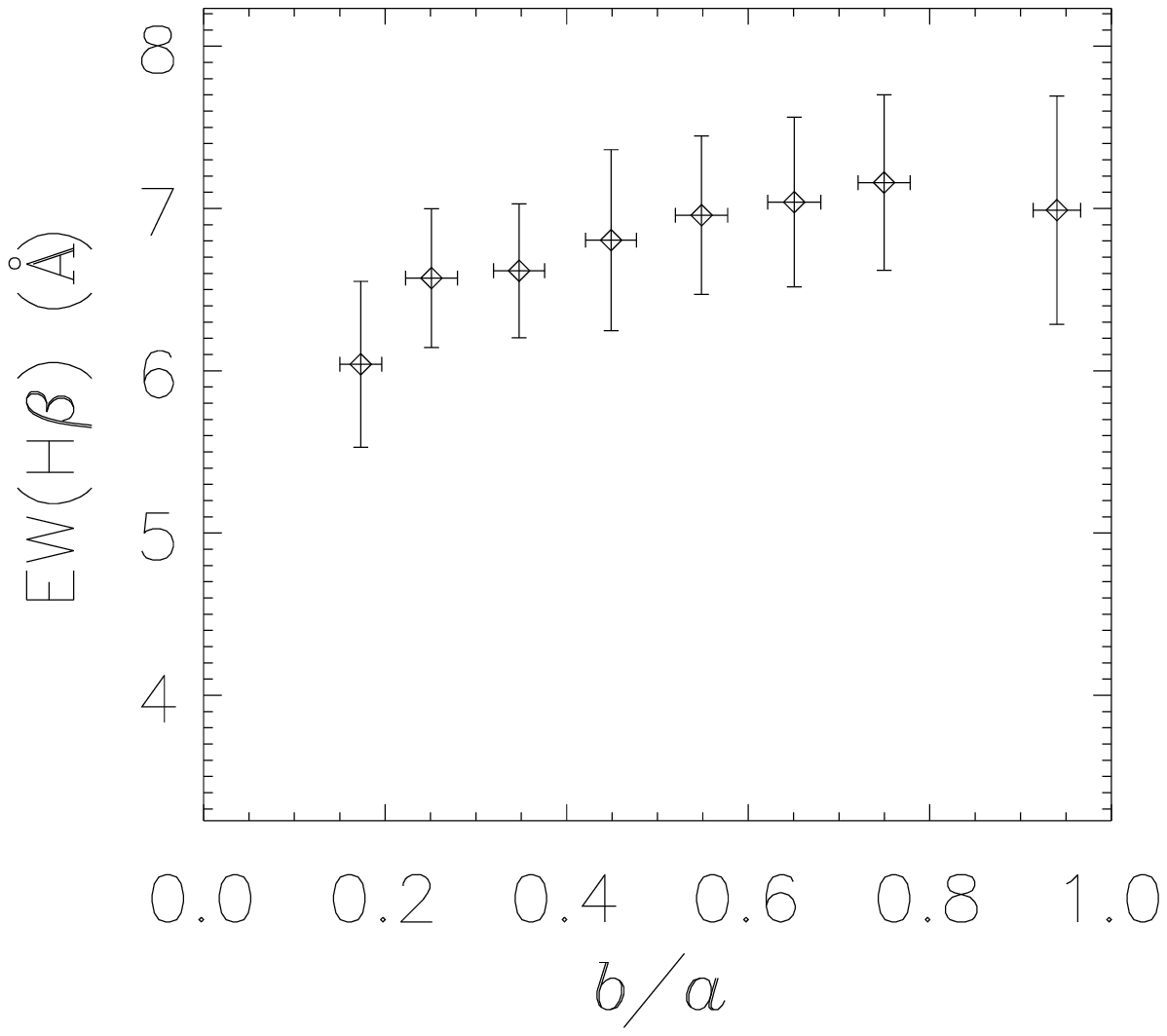}
\caption{The inclination dependency of  the \Hbeta \ equivalent width.
    The  sliding  window approach  (left)  and  the Bayesian  approach
    (right) are  used respectively in estimating the  continuum in the
    composite  spectra.  The  error  bar is  \onesigma  \  statistical
    uncertainty for each parameter.}
\label{fig:hb_ew_inclin}
\end{center}\end{figure}

The inclination dependency of the \Halpha \ and \Hbeta \ EW's is shown
respectively        in        \figname\ref{fig:ha_ew_inclin}       and
\ref{fig:hb_ew_inclin}, for the sliding window approach (left figures)
and the  Bayesian approach (right  figures).  The offset  between both
approaches   is    $\approx~{7}$~\AA   \   for    the   \Halpha,   and
$\approx~{4}$~\AA \  for the \Hbeta.   If the stellar  absorptions are
not  taken  into  account  as  in the  sliding  window  approach,  the
\balmerdec  \ would  be  mis-taken  to be  approximately  a factor  of
\balmerdecPercentileVSBayesian \ larger for our galaxies.

\begin{figure}\begin{center}
\epsscale{0.8}\plotone{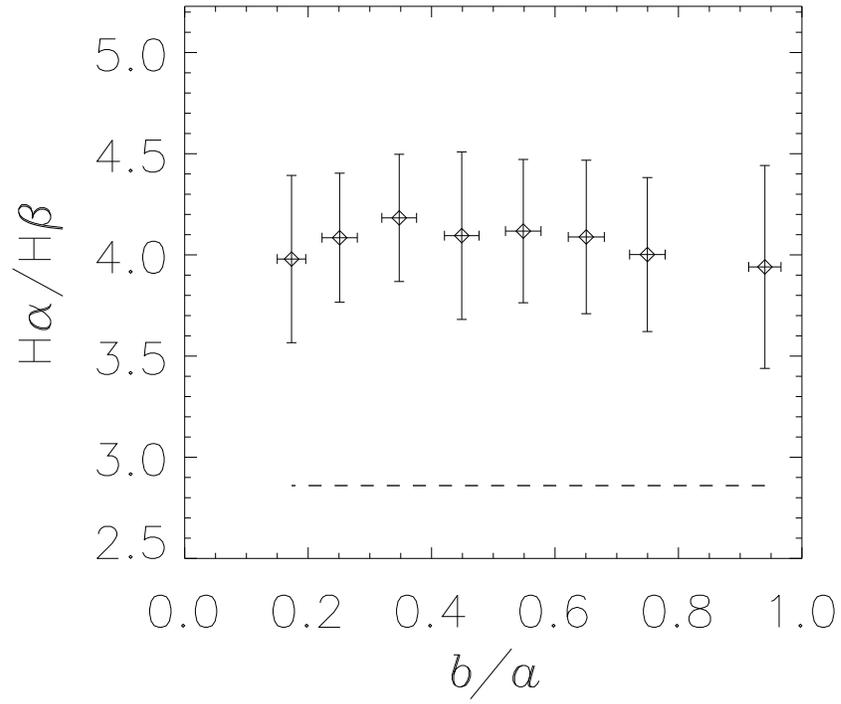}
\caption{The inclination  dependency of  the Balmer decrement  for the
  star-forming  disk  galaxies.   The  Bayesian approach  is  used  in
  estimating  the   stellar  continuum  in   the  geometric  composite
  spectra. The  dashed line indicates the theoretical  value of normal
  \ion{H}{2}  region.    No  obvious   trend  of  \balmerdec   \  with
  inclination is found. }
\label{fig:balmerdec_inclin.bayesian.gmean}
\end{center}\end{figure}

\begin{figure}\begin{center}
\epsscale{0.8}\plotone{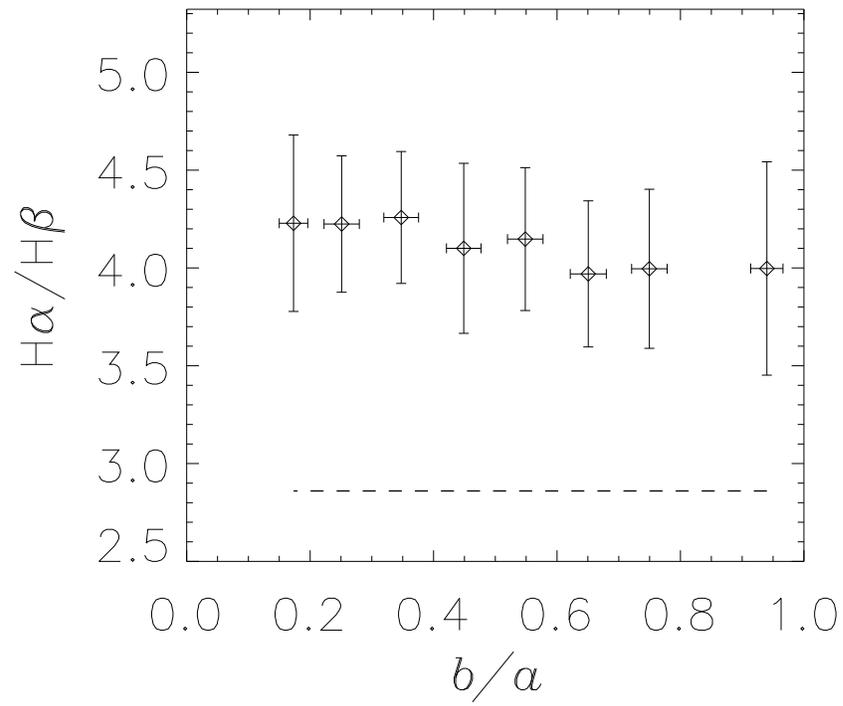}
\caption{Same as \figname\ref{fig:balmerdec_inclin.bayesian.gmean} but
using  the median  composite spectra.  The dashed  line  indicates the
theoretical value  of normal \ion{H}{2} region.   The Balmer decrement
values  are  consistent  to  \onesigma  \ uncertainty  with  those  in
\figname\ref{fig:balmerdec_inclin.bayesian.gmean}.}
\label{fig:balmerdec_inclin.bayesian.median}
\end{center}\end{figure}

The inclination  dependency of the  measured \balmerdec \ is  shown in
\figname\ref{fig:balmerdec_inclin.bayesian.gmean}, using the geometric
mean spectra.  The Balmer decrements are also calculated in the median
spectra,  shown in \figname\ref{fig:balmerdec_inclin.bayesian.median},
which   is  expected   to   preserve  the   relative  line   strengths
\citep{2001AJ....122..549V}.   The \balmerdec  \ calculated  using the
median composite  spectra are consistent with those  obtained by using
the geometric composite spectra.   Both plots show that the \balmerdec
\  remains fairly constant  with inclination  -- a  different behavior
from  the  stellar   continuum  extinction,  which  shows  inclination
dependency.

\subsection{Reddening in \ion{H}{2} Region vs. in Continuum-Generating Stars}

We         adopt        the        Balmer         optical        depth
\citep[$\balmertau$,][]{1994ApJ...429..582C},  a measure  of  the dust
reddening from the measured Balmer line ratio, as follows

\begin{equation}
  \balmertau              =              \log_{e}              {\left(
\frac{\Halpham/\Hbetam}{(\Halpham/\Hbetam)^{T}} \right) } \ .
\end{equation} 

\noindent
The  superscript  $l$  was  used  by  the  authors  to  indicate  that
$\balmertau$  derived from the  emission lines  \Halpha \  and \Hbeta.
The   ratio  $(\Halpham/\Hbetam)^{T}$   is   the  un-reddened   Balmer
decrement, taken to be the theoretical value, \balmerdecHIItheoretical
\ \citep{1986A&A...155..297C,1989agna.book.....O}  in this work, which
is accurate to 1~\% for $T_{e} = 5,000 - 20,000$~K.

\begin{figure}\begin{center}
\epsscale{0.8}\plotone{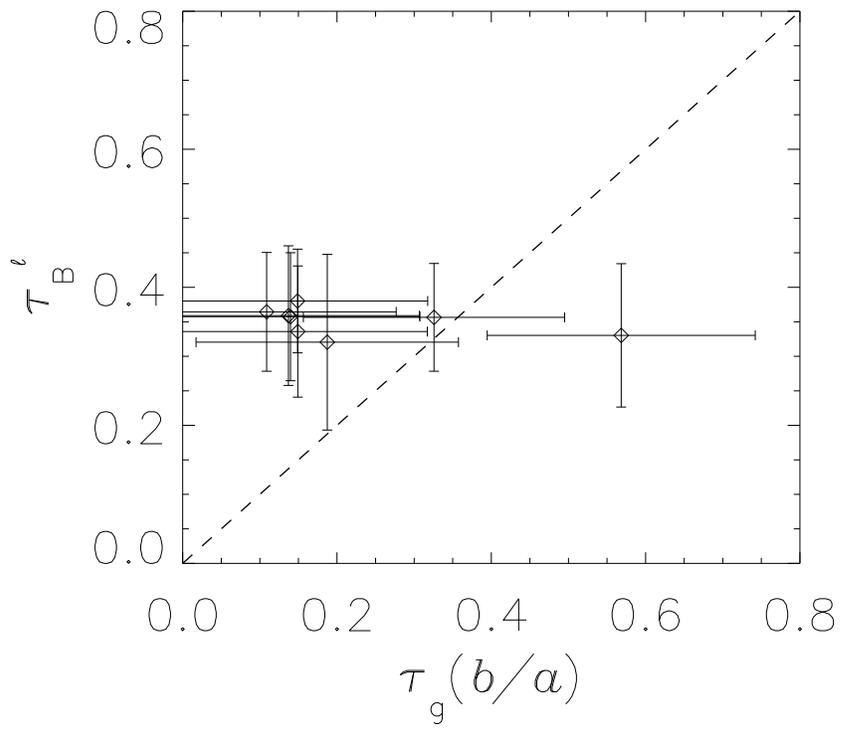}
\caption{The  Balmer  optical  depth  vs.  the  \sdssg-band  continuum
optical  depth.  The  error bar  in each  axis represents  \onesigma \
uncertainty in  the derived parameter.  The dashed  line indicates the
locus if both optical depths are the same. }
\label{fig:balmertau_abstau_g_bayesian}
\end{center}\end{figure}

The  Balmer optical depth  vs.  the  inclination-dependent \sdssg-band
stellar  continuum   optical  depth  ($\tau_{g}(\ba)$)   is  shown  in
\figname\ref{fig:balmertau_abstau_g_bayesian}.   The  face-on  optical
depths  ($\tau_{g}(1)$) are  taken to  be the  best-fit values  in the
\modelofchoice   \  (Table~\ref{tab:extfaceon}).    One   relation  is
evident, that  most of the points  of small optical  depths deviate by
more than \onesigma \  sample scatter from the one-to-one relationship
(the  dashed  line  in  \figname\ref{fig:balmertau_abstau_g_bayesian},
where $\balmertau=\tau_{g}(\ba)$), because $\balmertau$ is relatively
larger. 

Next, the color excess obtained  from the Balmer decrement is compared
with that derived from the inclination-dependent continuum extinction.
The intrinsic color excess  derived from the Balmer decrement, labeled
as  $\colorexcessbalmer$ \  in this  work,  is related  to the  Balmer
optical depth as follows \citep[][]{1994ApJ...429..582C}

\begin{equation}
\colorexcessbalmer \approx 0.935 \, \balmertau \ . 
\end{equation}

\noindent
 The   color  excess  in     stars,   $\colorexcessstar^{\ba}$,  is
calculated at a given inclination, $\ba$, by using

\begin{eqnarray}
\colorexcessstar^{\ba} & = & \BVstar^{\ba}  - \BVstar^{\rm 0} \\ & = &
                       A_{B}(\ba) - A_{V}(\ba) \ ,
                       \label{eqn:colorexcessstars}
\end{eqnarray}

\noindent
where $\BVstar^{\rm  0}$ is the intrinsic stellar  Johnson $B-V$ color
of the  stars, and  $\BVstar^{\ba}$ is the  reddened color due  to the
inclination effect.  Looking at \eqnname\ref{eqn:colorexcessstars} and
\eqnname\ref{eqn:relext}, the value of $\colorexcessstar^{\ba}$ can be
calculated    by   using    the   empirical    continuum   extinctions
$A_{B}(\ba)-A_{B}(1)$ and $A_{V}(\ba)-A_{V}(1)$, obtained in a similar
fashion as  in \S\ref{section:extinction} by using  the Johnson filter
response curves. The face-on extinctions $A_{B}(1)$ and $A_{V}(1)$ are
taken     to     be     those     from    the     \modelofchoice     \
(Table~\ref{tab:extfaceon}).  The  $\colorexcessstar$ \ depends weakly
on the model, because  the model-dependency of the face-on extinctions
in  the two  photometric  bands  is reduced  in  the subtraction  when
calculating $\colorexcessstar$.  If the face-on  extinction derived by
the screen  model is used instead,  the points are to  be moved toward
left horizontally  by \ebvmodeldependency \  in the plot  (not shown),
which do not affect the above results qualitatively.

\begin{figure}\begin{center}
\epsscale{0.8}\plotone{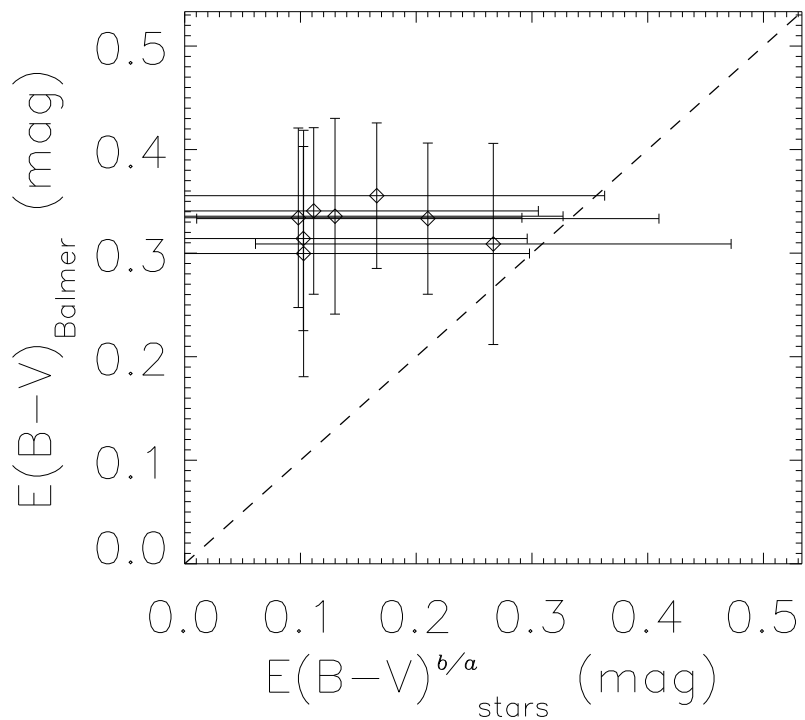}
\caption{The  comparison between  the  color excess  derived from  the
  Balmer  decrement and from  the inclination-dependent  extinction in
  the  stellar continuum, respectively.   The error  bar in  each axis
  represents \onesigma  \ uncertainty  in the derived  parameter.  The
  dashed  line indicates  the  locus  if every  galaxy  in the  sample
  exhibits uniform  interstellar extinction, where $\colorexcessbalmer
  = \colorexcessstar$ at a given inclination.}
\label{fig:ebv_balmer_vs_stars_mag_jhn_bayesian}
\end{center}\end{figure}

The  comparison between  the Balmer-line  and stellar  continuum color
excess                  is                  plotted                 in
\figname\ref{fig:ebv_balmer_vs_stars_mag_jhn_bayesian}.   Most  of the
points deviate by more than \onesigma \ uncertainty from the ``uniform
interstellar extinction'' line \citep{1986A&A...155..297C}, the dashed
line  in  each sub-figure  which  indicates  the  presence of  uniform
interstellar  extinction in  every  galaxy of  the  sample.  And,  the
Balmer-line color excess is larger than the continuum color excess.

If   we   assume   the    color   excesses   are   related   so   that
$\colorexcessstar^{\ba}  = c  \, \colorexcessbalmer$,  where $c$  is a
constant, then we find

\begin{equation}
\colorexcessstar^{\ba}   =    \left(   \ebvstartobalmer   \right)   \,
\colorexcessbalmer
\end{equation}

\noindent
for  our  sample.   The  constant  is  calculated  using  $c  =  \sum
\colorexcessstar^{\ba} \, / \sum  \colorexcessbalmer$, over all of the
inclinations.        The       result      $\colorexcessbalmer       >
\colorexcessstar^{\ba}$    agrees   qualitatively    with    that   by
\cite{2000ApJ...533..682C}  (their   \eqnname3,  with  a  proportional
constant of $0.44 \pm  0.03$), although our proportional constant is
approximately  a factor  of 2  lower,  both values  are consistent  to
\onesigma    \   uncertainty.    We    discuss   these    results   in
\S\ref{section:NIE}.

\section{Discussion \& Summary} \label{section:discussion}

\subsection{Spatially Non-uniform Interstellar Extinction} \label{section:NIE}

Our study shows  the lack of a correlation  between the Balmer optical
depth  and the continuum  optical depth,  and between  the Balmer-line
color excess and  the continuum color excess.  If  every galaxy of the
sample  exhibits  uniform interstellar  extinction,  one would  expect
$\colorexcessbalmer = \colorexcessstar$ at a given inclination.  These
results are likely the  manifestations of the non-uniform interstellar
extinction in  at least some  of the galaxies  in each of  the sample.
Found also earlier  in the \ion{H}{2} regions in  the Large Magellanic
Cloud  \citep{1986A&A...155..297C}, it  describes the  situation where
the  optical thickness  of the  dust  in front  of the  light-emitting
regions is  not the same across  the area being observed  on a galaxy.
Combining with  the factor  in which the  spatial distribution  of the
star-forming \ion{H}{2} regions can be different (e.g., clumped and/or
patchy) from  that of the  continuum-generating stars (e.g.,  low mass
stars), the  measured extinctions in  the stars and in  the \ion{H}{2}
regions thus can be different.

\subsection{Summary} \label{section:summary}

Our  study  provides   a  determination  of  the  wavelength-dependent
extinction  in a large  and well-defined  sample of  star-forming disk
galaxies, independent from previous works in the literature.  We found
that both  the luminosity and the  steepness of the  spectral slope of
the composite spectra decrease  (i.e., extinguished and reddened) with
the inclination of disk galaxies, in a volume-limited sample in $z = $
\sampleredshift \ and  $M_r = $ \samplemagnitude \  from the SDSS DR5.
Assuming  this effect  is mainly  due to  intrinsic extinction  in the
galaxies,  in the SDSS  \sdssg-band the  inclined relative  to face-on
continuum extinctions are found empirically  to be \rextg \ for highly
inclined  objects  ($\ba =  $  \baedgeonextra).   The derived  face-on
extinction  is model  dependent, nonetheless  with both  the  slab and
sandwich  models giving  \extgfaceon  \  in the  \sdssg  \ band.   The
extinction curve shows  a variety of extinction, in  both the sign and
the amplitude,  for emission/absorption lines relative  to the stellar
continuum.  The  parameterized extinction curves  allow for correcting
the observed galaxy continuum against the inclination effect.

The \balmerdec  \ does not  show dependence with the  inclination, and
remain fairly constant. This  inclination dependency is different from
that  in the stellar  continuum, suggesting  the mechanism  and/or the
dust configuration responsible for  the reddening are different in the
\ion{H}{2} region  and in  the continuum-generating stars.   The color
excess $\colorexcessbalmer  > \colorexcessstar$ at  most inclinations,
rather than the equality of both.  From this we conclude that at least
some of  the galaxies  exhibit the spatially  non-uniform interstellar
extinction, possibly caused by the presence of the clumped dust in the
vicinity of the \ion{H}{2} region in the galaxies.

The next step is to  correct for the observed properties with existing
parameter-inclination       relations      \citep[e.g.,][and      this
work]{2007ApJ...659.1159S,2008ApJ...687..976U,2008ApJ...678L.101D,2009ApJ...691..394M},
re-select the  galaxies, and repeat the analysis.   The convergence of
the  parameter-inclination relations is  essential for  ensuring their
correctness.     This     approach    is    with     reference    with
\citet{2007MNRAS.379.1022D},  who developed  an iterative  approach to
derive cosmic  mass densities of  the bulge, disk and  dust components
due  to  disk  galaxies,  and found  converged  extinction-inclination
relations for those galaxies.

Our  study also  raises several  interesting questions.   For example,
does the variation of extinction as a function of inclination from our
current  empirical measurement  agree with  that by  other approaches,
such as stellar  population synthesis modeling?  Can we  build a model
to   describe  the  distributions   of  dust,   gas,  stars   and  
line-emitting  regions,  that would  explain  the  relative amount  of
extinction in the emission/absorption lines and the stellar continuum, to
meet the extinction curve?  Can we apply the current analysis to other
galaxy types?  What is the effect of galaxy inclination on photometric
redshift, given that the inclination  causes a reddening to the colors
of a  galaxy? We will leave  these studies to  future papers.  Because
our sample selection is well defined in a publicly available data set,
works can be carried out easily even by other researchers in exploring
any secondary factors that may affect the extinction in the galaxies.

\section{Acknowledgments}

CWY  thanks Reynier  F.   Peletier for  discussions  on extinction  in
galaxies.   We  thank  the  referee  for  informative  and  insightful
comments,  particularly  on  the  sample  selection.   We  acknowledge
support through grants  from the W.M.  Keck Foundation  and the Gordon
and Betty  Moore Foundation, to establish a  program of data-intensive
science at  to the  Johns Hopkins University.   IC and  LD acknowledge
support from NKTH:Polanyi and KCKHA005 grants.

This research has made use  of data obtained from or software provided
by  the US  National Virtual  Observatory, which  is sponsored  by the
National Science Foundation.

Funding for  the SDSS and SDSS-II  has been provided by  the Alfred P.
Sloan Foundation, the Participating Institutions, the National Science
Foundation, the  U.S.  Department of Energy,  the National Aeronautics
and Space Administration, the  Japanese Monbukagakusho, the Max Planck
Society,  and the Higher  Education Funding  Council for  England. The
SDSS Web Site is http://www.sdss.org/.

The SDSS is  managed by the Astrophysical Research  Consortium for the
Participating  Institutions. The  Participating  Institutions are  the
American Museum  of Natural History,  Astrophysical Institute Potsdam,
University  of Basel,  University of  Cambridge, Case  Western Reserve
University,  University of Chicago,  Drexel University,  Fermilab, the
Institute  for Advanced  Study, the  Japan Participation  Group, Johns
Hopkins University, the Joint  Institute for Nuclear Astrophysics, the
Kavli Institute  for Particle  Astrophysics and Cosmology,  the Korean
Scientist Group, the Chinese  Academy of Sciences (LAMOST), Los Alamos
National  Laboratory, the  Max-Planck-Institute for  Astronomy (MPIA),
the  Max-Planck-Institute  for Astrophysics  (MPA),  New Mexico  State
University,   Ohio  State   University,   University  of   Pittsburgh,
University  of  Portsmouth, Princeton  University,  the United  States
Naval Observatory, and the University of Washington.

{}

\clearpage

%

\clearpage

\clearpage

\begin{figure}\begin{center}
\epsscale{1.0}\plotone{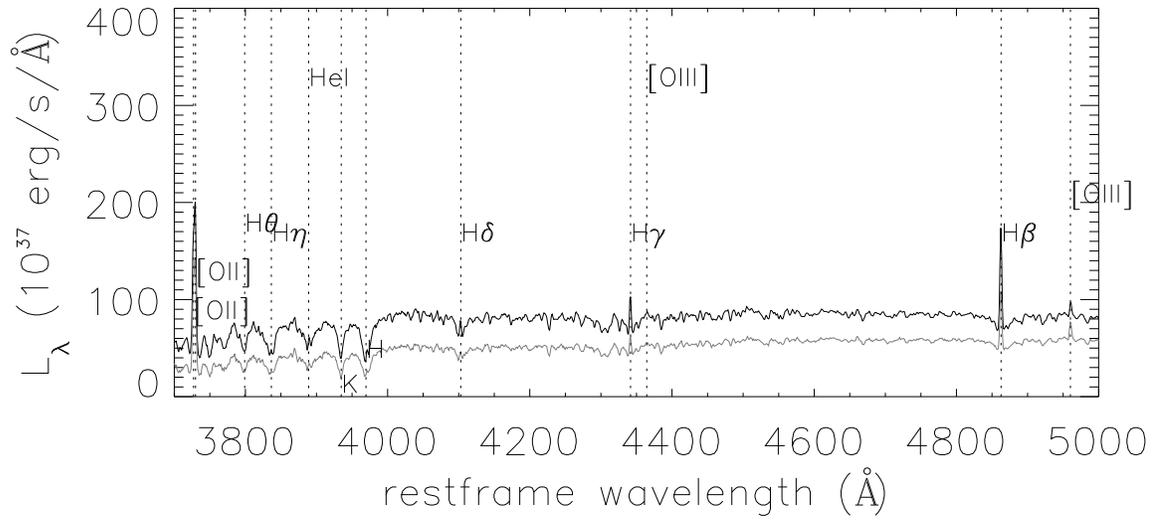}
\caption{The  comparison between  the restframe  composite  spectra of
  face-on ($\ba =  0.9 - 1.0$, black) and edge-on ($\ba  = 0.1 - 0.2$,
  gray)   star-forming   disk   galaxies,   for   the   volume-limited
  sub-sample. The flux density  per unit wavelength is an ``observed''
  quantity in the sense that inclination correction is not applied.}
\label{fig:flux_edgeon_faceon_0.1_0.2.gmean.3700._5000.}
\end{center}\end{figure}

\begin{figure}\begin{center}
\epsscale{1.0}\plotone{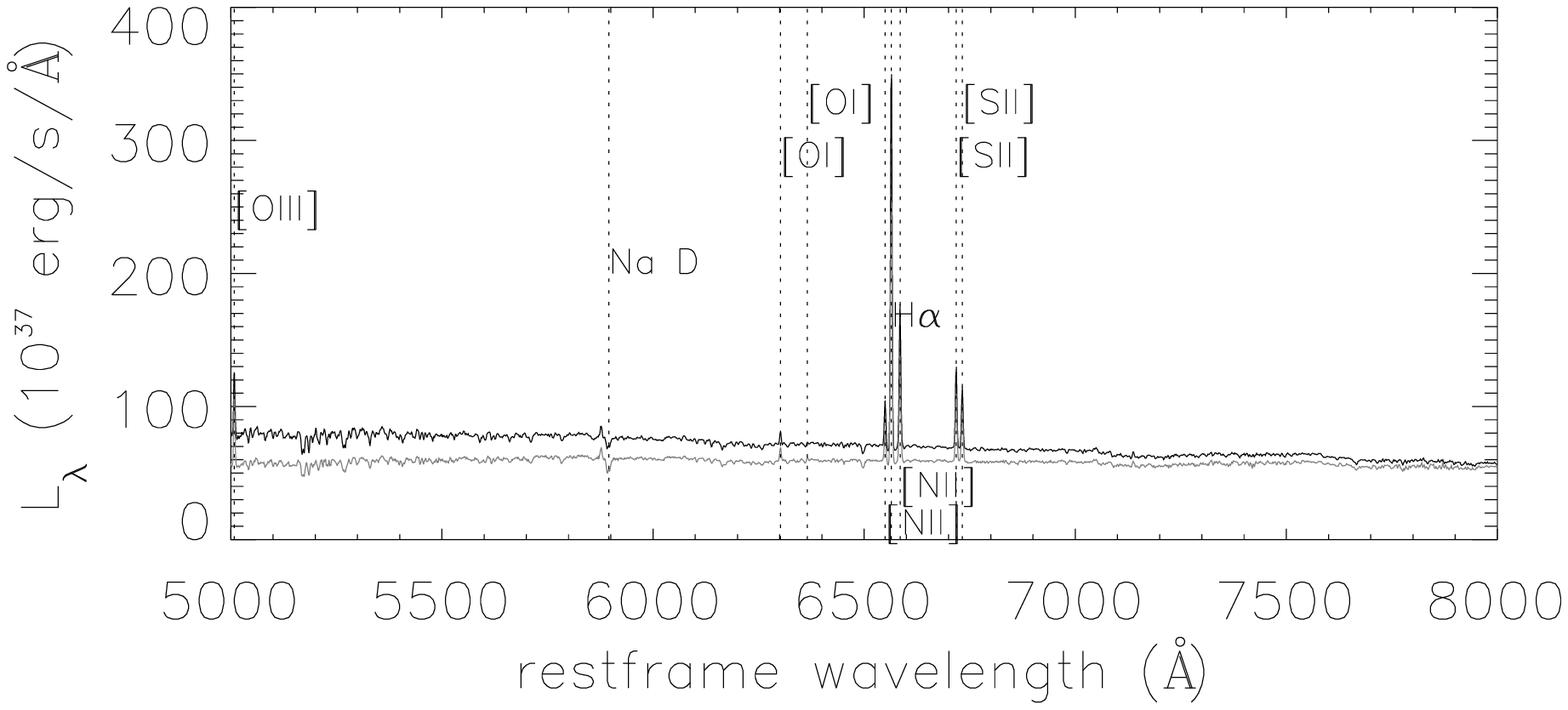}
\caption{Continuation                                                of
  \figname\ref{fig:flux_edgeon_faceon_0.1_0.2.gmean.3700._5000.}     to
  wavelengths $\waveMid - \waveE$~\AA.}
\label{fig:flux_edgeon_faceon_0.1_0.2.gmean.5000._8000.}
\end{center}\end{figure}

\clearpage

\clearpage

\begin{figure}\begin{center}
\epsscale{1.0}\plotone{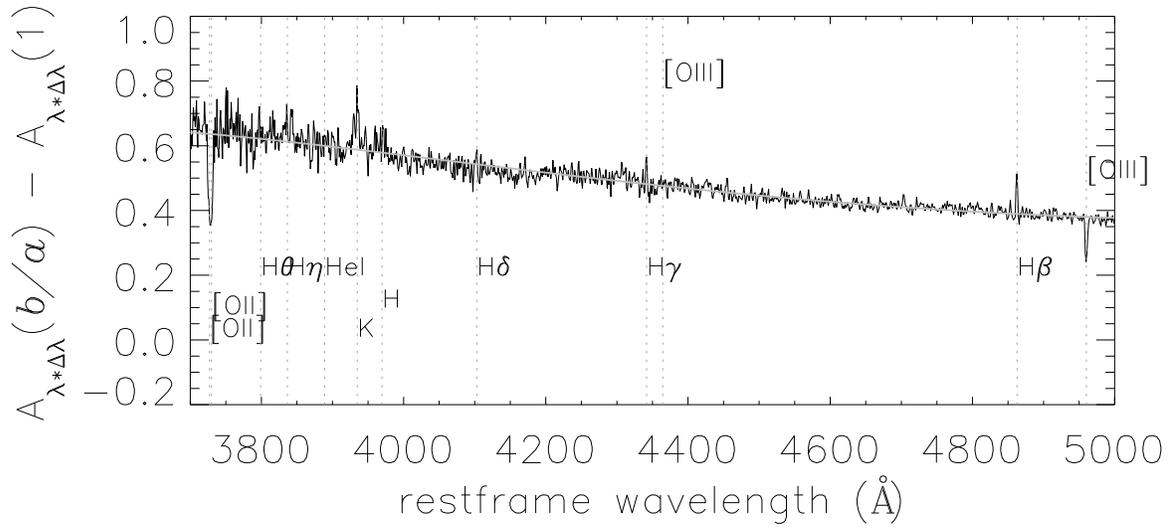}
\caption{The extinction curve obtained  by comparing the edge-on ($\ba
  = 0.1  - 0.2$)  with face-on  ($\ba = 0.9  - 1$)  composite spectra,
  according  to  \eqnname\ref{eqn:relext}.    The  gray  line  is  the
  best-fit  3rd  degree   polynomial  function  on  the  emission-line
  excluded                        continuum.                       See
  \figname\ref{fig:extin_0_0.3.ndeg.3.5000._8000.}    for    restframe
  wavelengths longer than 5000~\AA.}
\label{fig:extin_0_0.3.ndeg.3.3700._5000.}
\end{center}\end{figure}

\begin{figure}\begin{center}
\epsscale{1.0}\plotone{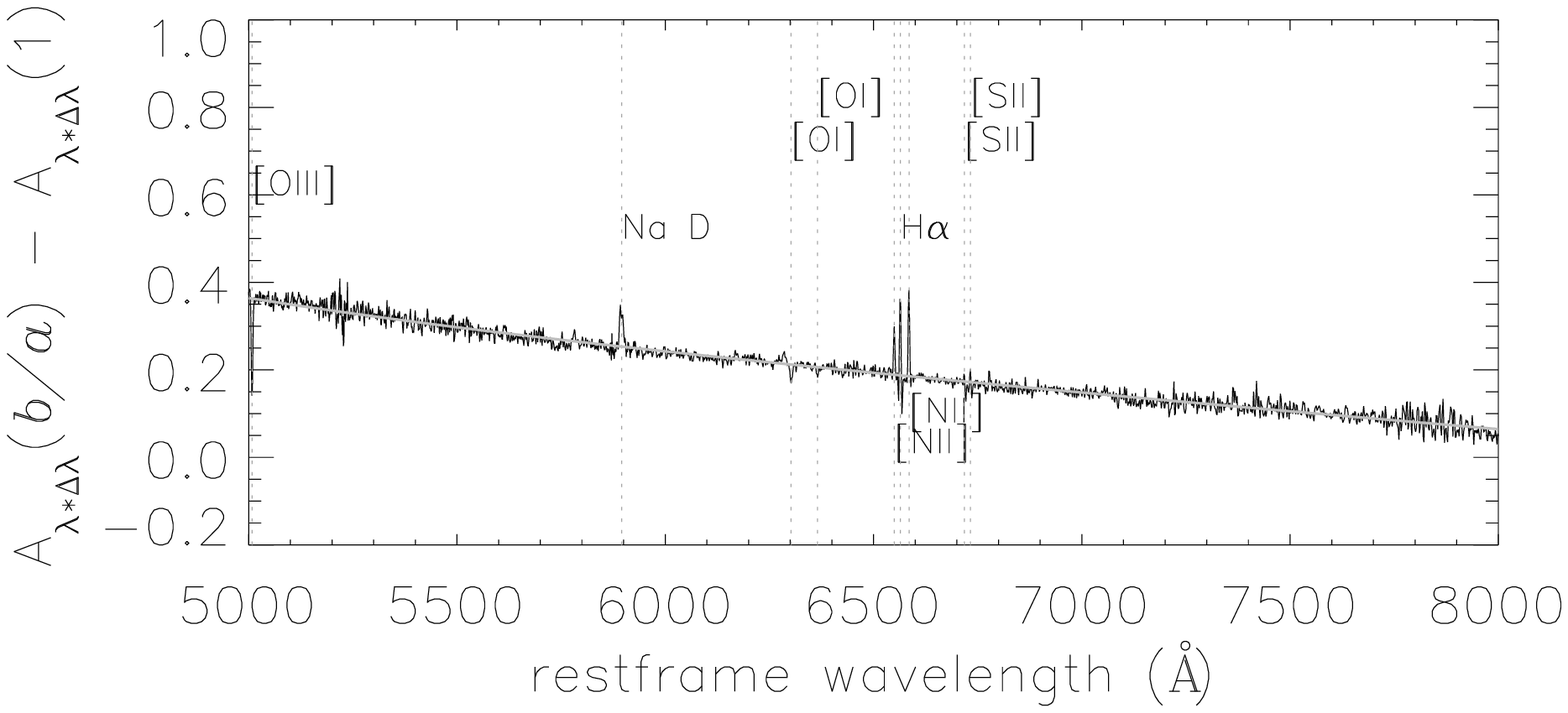}
\caption{Continuation of
  \figname\ref{fig:extin_0_0.3.ndeg.3.3700._5000.} to wavelengths
  $\waveMid - \waveE$~\AA.}
\label{fig:extin_0_0.3.ndeg.3.5000._8000.}
\end{center}\end{figure}

\clearpage

\clearpage

\clearpage

\clearpage

\clearpage

\begin{table}\begin{center}
\caption{Number of star-forming disk galaxy spectra in the volume-limited sample.}
\begin{tabular}{cc}
\hline 
$\ba$\tablenotemark{a} & number \\
\hline  
$0.94 \pm 0.03$ &     387
 \\
$0.75 \pm 0.03$ &     615
 \\
$0.65 \pm 0.03$ &     625
 \\
$0.55 \pm 0.03$ &     621
 \\
$0.45 \pm 0.03$ &     678
 \\
$0.35 \pm 0.03$ &     784
 \\
$0.25 \pm 0.03$ &     684
 \\
$0.17 \pm 0.02$ &     214
 \\
$0.09 \pm 0.00$ &     2
 \\
\hline 
\tablenotetext{a}{The average inclination ($\pm$ \onesigma \
  in the sample scatter) of the galaxies.} 
\end{tabular}
\label{tab:number}
\end{center}\end{table}

\begin{table}\begin{center}
\caption{Relative  extinction\tablenotemark{a} in disk galaxies as a function of
  inclination, in SDSS $g,r,i$ bands.}
\begin{tabular}{ccrrr}
\hline
\hline
{$\ba$}\tablenotemark{b} & 
  {$A_{g}(\ba) - A_{g}(1)$}\tablenotemark{c} &  {$A_{r}(\ba) - A_{r}(1)$} & {$A_{i}(\ba) - A_{i}(1)$} \\
\hline
$ 0.94\pm 0.03$
&$ 0.00\pm 0.18$
&$ 0.00\pm 0.12$
&$ 0.00\pm 0.15$
\\
$ 0.75\pm 0.03$
&$-0.04\pm 0.18$
&$-0.04\pm 0.12$
&$-0.04\pm 0.15$
\\
$ 0.65\pm 0.03$
&$-0.05\pm 0.18$
&$-0.05\pm 0.12$
&$-0.05\pm 0.15$
\\
$ 0.55\pm 0.03$
&$-0.09\pm 0.18$
&$-0.10\pm 0.12$
&$-0.11\pm 0.15$
\\
$ 0.45\pm 0.03$
&$-0.05\pm 0.18$
&$-0.09\pm 0.12$
&$-0.11\pm 0.15$
\\
$ 0.35\pm 0.03$
&$-0.04\pm 0.18$
&$-0.12\pm 0.12$
&$-0.16\pm 0.15$
\\
$ 0.25\pm 0.03$
&$ 0.15\pm 0.18$
&$ 0.02\pm 0.12$
&$-0.06\pm 0.15$
\\
$ 0.17\pm 0.02$
&$ 0.41\pm 0.19$
&$ 0.23\pm 0.13$
&$ 0.11\pm 0.15$
\\
\hline
\tablenotetext{a}{The restframe continuum relative extinction value,
  in magnitude. Emission lines
  are excluded by using the Bayesian approach.} 
\tablenotetext{b}{The average inclination ($\pm$ \onesigma \
  in the sample scatter) of the galaxies.} 
\tablenotetext{c}{The relative extinction ($\pm$ \onesigma \
  uncertainty) of the galaxies, in magnitude.} 
\end{tabular}
\label{tab:extrel}
\end{center}\end{table}

\begin{table}\begin{center}
\caption{Best-fit face-on extinction\tablenotemark{a} of star-forming
   disk galaxies in  SDSS \sdssg, \sdssr, \sdssi \ bands.}
\begin{tabular}{ccccccc}
\hline 
\hline  
model & 
$A_{g}(1)$\tablenotemark{b}  & reduced \chisq  &
 $A_{r}(1)$  & reduced \chisq  &      
$A_{i}(1)$  & reduced \chisq  \\
\hline 
  screen & $0.06 \pm 0.03$ & 0.36 & $0.02 \pm 0.02$ & 0.67 & $-0.01\pm0.02$ & 0.47\\
    slab & $0.20 \pm 0.11$ & 0.17 & $0.09 \pm 0.06$ & 0.47 & $0.02\pm0.07$ & 0.45\\
sandwich & $0.15 \pm 0.05$ & 0.19 & $0.14 \pm 0.14$ & 0.53 & $0.15\pm0.18$ & 0.45\\
\hline 
\end{tabular}
\label{tab:extfaceon}
\tablenotetext{a}{The restframe continuum extinction value. Emission lines
  are excluded by using the Bayesian approach.} 
\tablenotetext{b}{The best-fit face-on extinction ($\pm$ \onesigma \ uncertainty)
  of the galaxies, in magnitude.} 
\end{center}\end{table}

\begin{table}\begin{center}
\caption{Extinction curve\tablenotemark{a} in star-forming disk galaxies,
  parameterized with 3rd degree polynomial\tablenotemark{b}.}
\begin{tabular}{rrrrr}
\hline
\hline
$a_{0}$ & $a_{1}$ & $a_{2}$ & $a_{3}$ & reduced $\chi^2$ \\
\hline
$  -0.554 \pm    0.014 $ & 
$   0.564 \pm    0.023 $ & 
$  -0.057 \pm    0.012 $ & 
$   0.003 \pm    0.002 $ & 
  0.15 \\
\hline
\end{tabular}
\label{tab:extcurve}
\tablenotetext{a}{Obtained by comparing between the edge-on ($\ba = \baedgeonrange$)
  and face-on ($\ba = \bafaceonrange$) composite spectra in the restframe
  wavelengths $\waveS -\waveE$~\AA. Emission lines are
  excluded by using the Bayesian approach.} 
\tablenotetext{b}{$A_{\lambda * \Delta\lambda}(\ba = \baedgeonrange)  -   A_{\lambda * \Delta\lambda}(1)  =  \sum_{j  =
  0}^{3}  \, a_{j}  \, {\wavenumber}^{j}$,  where  the inverse
  restframe wavelength ${\wavenumber}$ is in the unit of $\micron^{-1}$.} 
\end{center}\end{table}

\end{document}